\documentclass[aps,pre,twocolumn,showpacs,superscriptaddress,groupedaddress,amsmath,amssymb,floatfix]{revtex4-1} 
\usepackage{bm}
\usepackage{float}
\usepackage{amsfonts}
\usepackage{times}
\usepackage[dvips]{graphicx}
\usepackage{color}
\usepackage{ulem}
\usepackage{chemist}
\usepackage{lipsum}
 \usepackage[utf8]{inputenc}

\begin{document}

\title{{Experimental study of three-wave interactions among capillary-gravity surface waves}}
\author{Florence Haudin$^1$}
\author{Annette Cazaubiel$^{1,2}$} 
\author{Luc Deike$^3$}
\author{Timoth\'ee Jamin$^1$} 
\author{Eric Falcon$^1$} 
\author{Michael Berhanu$^1$} 
\affiliation{$^1$Universit\'e Paris Diderot, Sorbonne Paris Cit\'e, MSC, CNRS (UMR 7057), 75013 Paris, France}
\affiliation{$^2$Ecole Normale Sup\'erieure, D\'epartement de Physique, 75005 Paris, France}
\affiliation{$^3$Scripps Institution of Oceanography, University of California San Diego, USA}

\date{February 27, 2016}

\begin{abstract}
In propagating wave systems, three or four-wave resonant interactions constitute a classical non-linear mechanism exchanging energy between the different scales. Here we investigate three-wave interactions for gravity-capillary surface waves in a closed laboratory tank. We generate two crossing wave-trains and we study their interaction. Using two optical methods, a local one (Laser Doppler Vibrometry) and a spatio-temporal one (Diffusive Light Photography), a third wave of smaller amplitude is detected, verifying the three-wave resonance conditions in frequency and in wavenumber. Furthermore, by focusing on the stationary regime and by taking into account viscous dissipation, we directly estimate the growth rate of the resonant mode. The latter is then compared to the predictions of the weakly non-linear triadic resonance interaction theory. The obtained results confirm qualitatively and extend previous experimental results obtained only for collinear wave-trains. Finally, we discuss the relevance of three-wave interaction mechanisms in recent experiments studying gravity-capillary turbulence.

\end{abstract}

   \newpage                          
\maketitle
\section{Introduction}
When a wave-field  is governed by linear propagation equations, the different space and time scales coupled by the dispersion relation evolve independently. In contrast, by introducing non-linearity, energy exchanges between the scales become possible. Among non-linear interaction phenomena, a particular attention has been given to the resonant interaction mechanisms ~\cite{Phillips1981,Craik,Hammack1993}. In the weakly non-linear limit, dispersive waves can substantially exchange energy at long time, if their respective angular frequencies $\omega_i= 2 \pi\, f_i$ and wavenumbers $\mathbf{k_i} \, \, \left( \Vert \mathbf{k_i} \Vert=k_i=\frac{2 \pi}{\lambda_i}\right) $ satisfy the resonance conditions. For a process involving $N$ waves, these conditions take the form:
 \begin{eqnarray}
\mathbf{k_1} \pm \mathbf{k_2} + \ldots \pm \mathbf{k_N} =0
\label{Eq1a}
\end{eqnarray}
\noindent
 \begin{eqnarray}
\omega (\mathbf{k_1}) \pm \omega (\mathbf{k_2}) + \ldots \pm \omega (\mathbf{k_N})=0
\label{Eq1b}
\end{eqnarray}
with $\omega (\mathbf{k_{i}})$ the dispersion relation of the waves.\\
At the lowest order of the weakly non-linear expansion, three-wave mechanism or triadic resonance is usually considered for a system with a quadratic non-linearity. Nevertheless if the dispersion relation cannot satisfy the three-wave resonance condition, four-wave mechanism occurs. This is precisely the case of large scale surface gravity waves~\cite{Phillips1960,Longuet-Higgins1962}, in which resonant interactions rule dynamics and evolution of oceanic gravity waves~\cite{Janssen2004}. At smaller scales, in the crossover between gravity and capillary waves and also for pure capillary waves, three-wave interactions occur~\cite{McGoldrick1965}. Triadic resonance is also involved in numerous natural and physical examples in which they mediate energy exchanges between the different scales, like internal gravity waves~\cite{Ball1964} in oceanography or for three-wave mixing in non-linear optics~\cite{Boydbook}. Recent experiments have indeed investigated triadic resonance and verified the resonance condition in internal waves~\cite{Joubaud2012}, in inertial waves in rotating tank \cite{Bordes2012}, in plasma waves \cite{Sokolov2014} or in hydro-elastic waves \cite{Deike2015}. 

\indent Moreover understanding resonant interaction mechanisms is of prime interest in the study of wave turbulence regimes. The dynamics of a set of numerous random waves in interaction, is said in a wave turbulence regime, when a self-similar mechanism transfers energy from an injection scale to a dissipative scale. The aim of the weak turbulence theory~\cite{Zakharovbook,Nazarenko2011} is to describe this regime, by taking the main wave resonant mechanism as the elemental process coupling the waves. Power spectra of wave amplitude can be analytically derived as power laws of $k$ (spatial spectrum) or $\omega$ (temporal spectrum). Recently several laboratory investigations of wave turbulence have been carried out in regard of weak turbulence theory, specifically for the example of hydrodynamics surface waves. At large scale, in the gravity wave regime, involving four-wave interactions, power spectra of wave elevation are generally in partial agreement with theoretical predictions. Exponent of the scale-power law depends on forcing amplitude and seems to saturate close to the predicted value~\cite{,NazarenkoJFM2010,DeikeJFM2015,Falcon2007}. At small scale, in the capillary wave regime, several independent studies report in contrast observation of the exponents given by the weak turbulence theory~\cite{Putterman1996,Henry2000,Brazhnikov2002,Falcon2007,FalconFalcon2009,Xia2010}. More recently the spatio-temporal characterization of capillary wave turbulence~\cite{Berhanu2013} and the study of its decay~\cite{Deike2012} have shown, that despite the compatibility of the spectra with weak turbulence theory, the needed experimental conditions depart from the theoretical framework. First viscous dissipation occurs at all the scales of the turbulent cascade leading to a non conserved energy flux~\cite{Deike2014} and anisotropy in the forcing is conserved~\cite{Berhanu2013}. Then the dimensionless non-linear parameter, \textit{i.e.} the wave steepness seems too large to verify the hypothesis of small non-linearity, needed to consider resonant interactions. Therefore in order to solve this paradox, an experimental investigation of three-wave resonant interactions of gravity-capillary surface waves is here performed, with similar experimental conditions than those exhibiting capillary wave turbulence. 

\indent Although this phenomenon has been widely studied, some important features have never been tested experimentally. Theoretically, by considering three waves verifying the resonant condition, amplitude equations can be derived using perturbative~\cite{McGoldrick1965,Case1977} or variational~\cite{Simmons1969} methods, in which dissipation is neglected. A few experimental studies tried to verify these last results in laboratory channel tanks, by generating a well controlled sinusoidal wave-train. The first investigations of such interactions were performed in the seventies for a special degenerate case, the Wilton ripples ($f_1=f_2=9.8$ Hz, $f_3=19.6$ Hz for pure water), with in that case the daughter wave collinear to the two mother waves~\cite{McGoldrick1970}. By adding dissipation as a perturbation, a conclusive agreement was found with theory. Similar experiments observed the same phenomenon under near-resonant conditions~\cite{Kim1971}. Then subharmonic generation of gravity-capillary waves were also studied at the beginning of the eighties~\cite{Banerjee1982}, in which two waves are generated from a single one at higher frequency. However, this three-wave process was not properly observed due to occurrence of subharmonic cross-waves~\cite{Hogan1984,Hammack1993}. At the end of the eighties, the subharmonic three-wave interaction phenomenon was appropriately reported~\cite{Henderson1987_1, Henderson1987_2, Henderson1987_3}, by demonstrating the instability of a capillary surface wave-train, whose frequency is above $19.6\,$Hz. The selection process of the observed triads was also studied.  Moreover, in a wave turbulence regime, a recent experimental study close to the gravito-capillary crossover has shown significant occurrence  of quasi 1D-three-wave interactions~\cite{Aubourg2015}. Finally, occurrence of three-wave interactions was demonstrated for capillary waves under parametric excitation~\cite{Punzmann2009}.

\indent In that context, the present paper investigates interactions between two gravity-capillary mother waves and a third daughter wave. By studying experimentally interactions between two sinusoidal wave trains producing a daughter wave at higher frequency, we intend to characterize the elemental process of wave turbulence, producing a positive transfer of energy through the small scales. This regime is indeed formed by a dynamic superposition of multiple interactions. In particular we aim to test the robustness of resonant interaction mechanisms knowing that in small scale experiment viscous dissipation is significant, multiple reflections occur, and time scale separation is not guaranteed allowing observation of non-resonant interactions~\cite{Kim1971,Janssen2004}. Moreover three-wave interaction mechanisms for surface waves have never been addressed experimentally in the configuration where two mother waves generate one daughter wave at higher frequency. Only the degenerated and collinear case~\cite{McGoldrick1970,Kim1971} ($f_1=f_2$; $f_3=2f_1$) and the subharmonic generation of two waves by one wave at high frequency~\citep{Banerjee1982,Hogan1984,Henderson1987_1} ($f_1=2f^*$; $f_2=f_3=f^*$) have been experimentally investigated.
 
\indent The article is organized as follows.  A state of the art is given in section I on the topics of gravity-capillary wave interactions whereas section II recalls the key points of the theoretical background of the resonant triadic interactions. Section III presents the experimental set up and techniques. Section IV reports an extensive study with local and spatial measurements of the triad, where two mother waves of frequency $f_1=15\,$ and $f_2=18$\,Hz generate a daughter wave at the frequency $f_3=f_1+f_2=33\,$Hz. Section V shortly extend the results to another triad experimentally tested ($f_1=16$, $f_2=23$, $f_3=39$ Hz) and draws some conclusions. Finally, in the Appendix, characterization of wave dissipation in the experimental set up is given.

\section{Three-wave resonance: theoretical background}
\subsection{Resonance conditions}
Let us consider the case of two mother waves with frequencies $f_1$ and $f_2$  and wavevectors $\mathbf{k_1}$ and $\mathbf{k_2}$, leading to the appearance of a daughter wave with frequency $f_3=f_1 + f_2$.
\noindent
 Each wave $i$ satisfies the gravity-capillary linear dispersion relation:
\begin{equation}
 \omega_i^2= \left[  g\,k_i+\frac{\sigma}{\rho}\, k_i ^3\right] \, \mathrm{tanh}(k _i\,h_0)
\label{disp}
\end{equation}
\noindent
with $g$ the gravity acceleration, $\rho$  the fluid density, $\sigma$ the surface tension and $h_0$ the liquid depth at rest. As the frequencies are imposed, norms of wave vectors $k _i$ are known by inverting numerically the relation dispersion. The components of the triad satisfy the resonance conditions:
\begin{eqnarray}
\omega_1+\omega_2 = \omega_3 &\hspace{3cm} &  \mathbf{k_1}+\mathbf{k_2}=\mathbf{k_3}
\label{Res}
\end{eqnarray}

\begin{figure}
\begin{center}
\includegraphics[height=0.5\columnwidth]{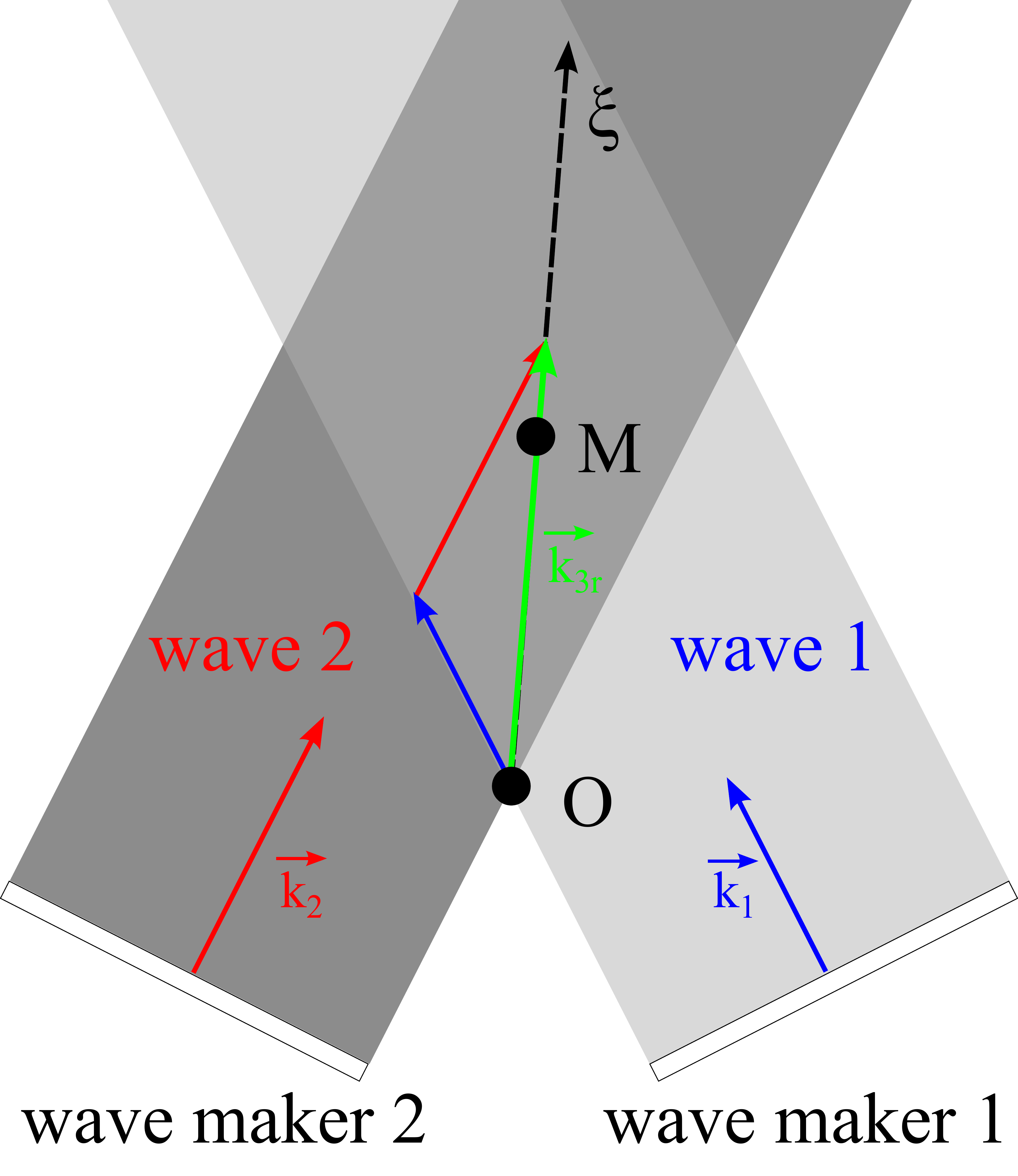}\hfill
\includegraphics[height=0.5\columnwidth]{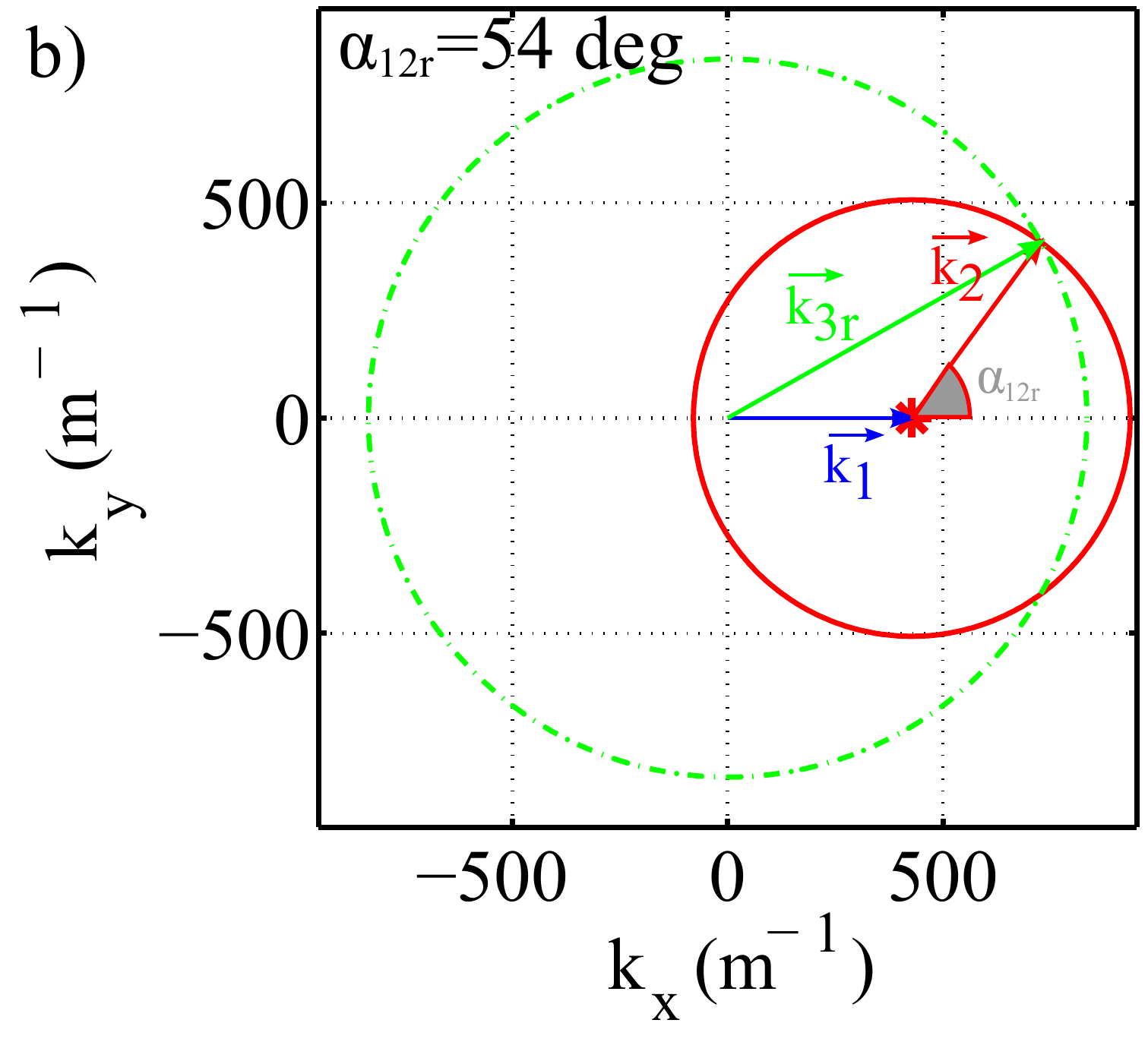}
\caption{(color online) a) Schematic view of the interaction zone between the two mother waves. $O$ is the origin of this zone, $M$ locates a point in this area along the direction $O_\xi$ given by $\mathbf{k_1}+\mathbf{k_2}$. b) Graphical construction of the triad for $f_1=15$ Hz and $f_2=18$ Hz with $\alpha_{12\,r}=54$ deg (resonant angle). The green circle (dash-dotted line) corresponds to $\mathbf{k_3r}$ satisfying the gravity-capillary dispersion relation and the red circle (continuous line) to all location of the extremity of vector $\mathbf{k_{2}}$. Axis are here defined relatively to $\mathbf{k_1}$.} \label{triad_graphs1}
\end{center}
\end{figure}

\begin{figure}[h!]
\begin{center}

\includegraphics[height=.60\columnwidth]{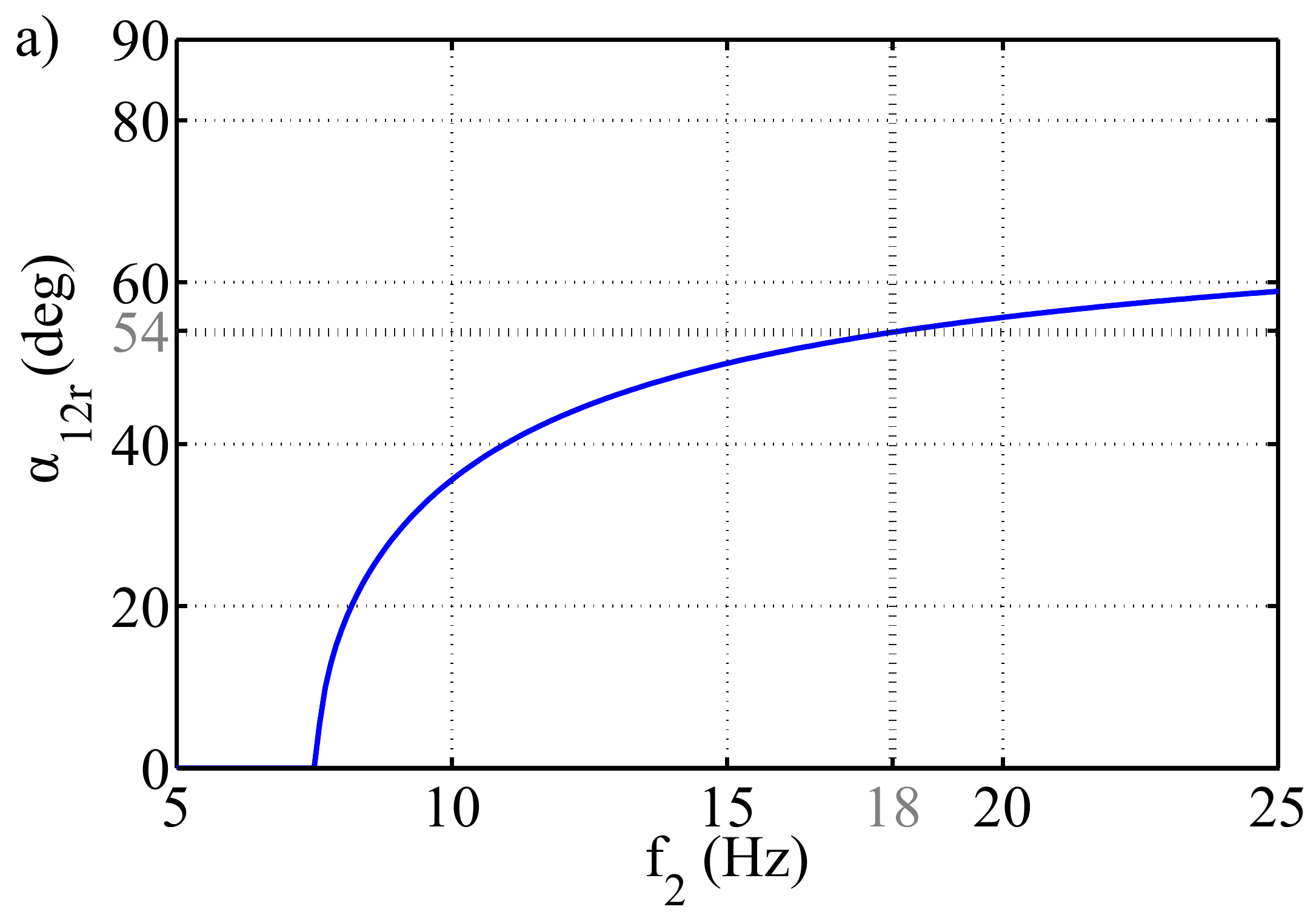}\hfill
\includegraphics[height=.63\columnwidth]{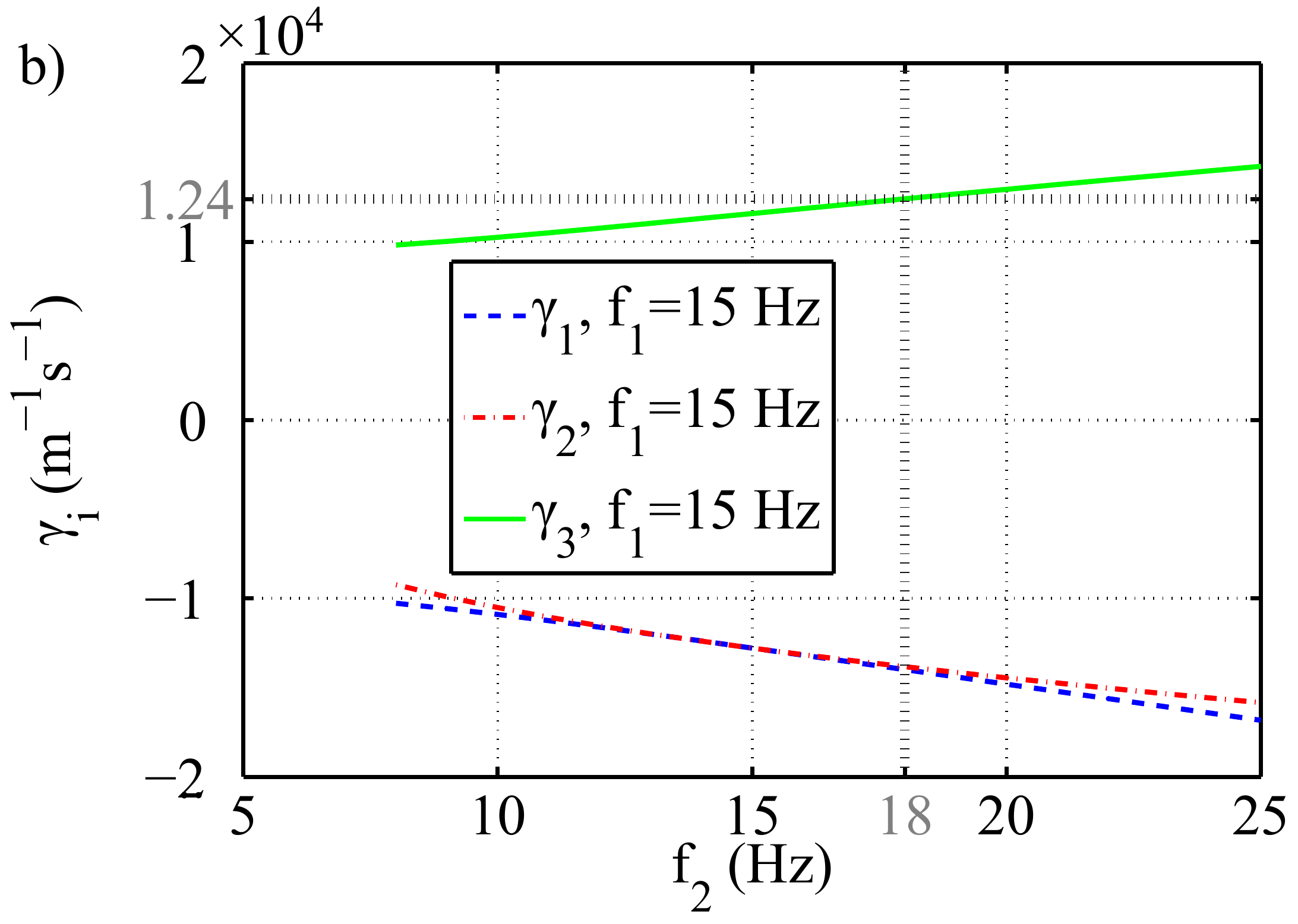} 

\caption{(color online) For a fixed mother wave $f_1=15$ Hz, a) evolution of the angle $\alpha_{12\,r}$ versus $f_2$ from Eq.~\ref{Eqalpha} and b) evolution of the interaction coefficients versus $f_2$ from Eq.~\ref{gamma}.}
\label{gammai1} 
\end{center}
\end{figure}

\noindent 
The angle between the two mother wave-vectors $\mathbf{k_1}$ and $\mathbf{k_2}$ is called $\alpha_{12}$. In the resonance conditions (Eq.~\ref{Res}), this angle noted $\alpha_{12\,r}$ is completely determined by the choice of mother wave frequencies $f_1$ and $f_2$.
\begin{equation}
\mathrm{cos}(\alpha_{12\,r})=\frac{k_3^2-(k_1^2+k_2^2)}{2 k_1 k_2 }
\label{Eqalpha}
\end{equation}
\noindent For the triads, we have investigated in the following, the values of $\alpha_{12\,r}$ are 54 deg  ($f_1=15$, $f_2=18$ and $f_3=33$ Hz) and 59 deg ($f_1=16$, $f_2=23$ and $f_3=39$ Hz), by taking $\sigma=60$ mN/m and $\rho=1000$ kg/m$^3$ for water as working fluid. These frequencies belong to the capillary waves domain, but gravity is not negligible and the complete form of the dispersion relation of Eq.~\ref{disp} has to be used. A schematic view of the wave beams of the mother waves and the resulting interaction zone area are given in Fig.\ref{triad_graphs1} a), with the corresponding angle $\alpha_{12\,r}$ between $\mathbf{k_1}$ and $\mathbf{k_2}$. An origin $O$ and an axis $O_\xi$ in the direction of $\mathbf{k_1}+\mathbf{k_2}$ are defined to describe the daughter wave propagation at a given point $M$ along $O_\xi$.\\
It is also possible to determine graphically the resonant wave vector $\mathbf{k_{3r}}$ satisfying both the dispersion relation and the resonance conditions as illustrated in Fig.~\ref{triad_graphs1} b). The red circle (continuous line) defines the loci of all the possible $\mathbf{k_3}$ built by the sum of $\mathbf{k_1}+\mathbf{k_2}$, different angles between them being possible when changing the orientation of the vector $\mathbf{k_2}$ on this circle, keeping its norm $k_2$ constant. The green circle (dash-dotted line) corresponds to the loci of the vectors $\mathbf{k_{3r}}$ in accordance with the dispersion relation. As a consequence the intersection between the red and the green circles defines the vector $\mathbf{k_{3r}}$ satisfying both the resonance conditions and the relation dispersion. Two solutions exist corresponding to opposite values of $\alpha_{12\,r}$. 
 If the angle between $\mathbf{k_1}$ and $\mathbf{k_2}$ differs from $\alpha_{12\,r}$, being for example $90$ deg, there is no intersection with the green circle and hence the corresponding triad is not satisfying the dispersion relation anymore. Moreover, varying $f_2$ for a given $f_1$ changes the value of $\alpha_{12\,r}$ computed with Eq.~\ref{Eqalpha}. Figure~\ref{gammai1} a) shows indeed a significant increase. At the lowest possible value of $f_2$,  $\alpha_{12\,r}=0$ and the mother waves are collinear. The value $\alpha_{12\,r}=90$ deg, cannot be reached and corresponds to $f_2$ going to infinity. Note also, that another graphical determination of triads is provided by the construction of Simmons~\cite{Simmons1969}, which builds in the 3D-space ($k_x$, $k_y$, $\omega$) the loci of the vector $k_i$ to be in the resonant situation. 

\subsection{Amplitude equations}
\label{Amplitudes}
In the weakly non-linear regime, considering one isolated triad, equations can be derived for the amplitudes and the phases of the three waves~\cite{Simmons1969}. By hypothesis, the free surface elevation $\eta$ is written as the sum of the components of the triad taken as propagative waves:\\

 $\eta(x,y,t)=\sum \limits_{i=1}^3\,a_i\,\cos (\mathbf{k_i} \cdot \mathbf{x} - \omega_i\,t +\phi_i)$, where $a_i$  and $\phi_i$ are the amplitude and the phase of the wave $i$. Temporal evolutions of $a_i$  and $\phi_i$ are then expressed as a system of six coupled first order linear differential equations. Moreover in capillary wave experiments, wave dissipation due to viscosity, needs to be taken into account. Considering a linear dissipation process, amplitude equations are modified by adding a decay term $\delta_i\,a_i$~\cite{McGoldrick1970,Henderson1987_1}, where $\delta_i$ is the viscous dissipation coefficient, \textit{i.e.} the inverse of the viscous decay time at the frequency $f_i$. This approach is justified, if this later time scale is small compared to the characteristic non-linear time associated with the non-linear growth of $a_3$. Then for each component $i$ of the triad, one has~\cite{Henderson1987_1}:
 
\begin{eqnarray}
\partial a_i/\partial t+\delta_i a_i+(\mathbf{v}_{gi}.\mathbf{\nabla})a_i &= & a_{i+1}a_{i+2} \, \gamma_i \sin\phi \label{EqAmp}\\
\partial \phi_i/\partial t+(\mathbf{v}_{gi}.\mathbf{\nabla})\phi_i & =& \dfrac{a_{i+1}a_{i+2}}{a_i} \, \gamma_i \cos\phi
\label{EqPhase}
\end{eqnarray}
\noindent 
with $i=1,2,3$ interchanged cyclically, $\phi$ the total phase defined as $\phi=\phi_1+\phi_2-\phi_3$, $\mathbf{\nabla}=(\partial_x,\partial_y)$, $\mathbf{v}_{gi}=\partial{\omega_i}/{\partial \mathbf{k_i}}$ group velocities and $\gamma_i$ the interaction coefficients, whose expressions are given by~ \cite{Simmons1969}:

\begin{equation}
\gamma_i= -\frac{k_i}{4\omega_i}\sum \limits_{j=1}^3 \, \omega_j \omega_{j+1}\left(1+\frac{\mathbf{k_j}.\mathbf{k_{j+1}}}{k_j k_{j+1}}\right)
\label{gamma}
\end{equation}
\noindent
In this last formula, the convention of negative frequency for $\omega_3$ is taken, in order to write the resonance conditions: $\omega_1+\omega_2+\omega_3=0$ and $\mathbf{k_1}+\mathbf{k_2}+\mathbf{k_3}=0$, as defined in the variational derivation of interaction coefficients~\cite{Simmons1969}. In the following except in the calculation of these coefficients, frequencies are considered positive.\\
\noindent
In order to provide some orders of magnitude, the evolution of the various $\gamma_i$ versus $f_2$ have been plotted on Fig.~\ref{gammai1} b), for a fixed value of $f_1=15$ Hz. The evolution with $f_2$ is monotonous (decreasing for $\gamma_{1,2}$ increasing for $\gamma_{3}$). The values of $\gamma_i$ for (15, 18, 33) and (16, 23, 39) (shortly discussed in section V) are reported in table I. 
\small

\begin{table}[h!]
\begin{minipage}[t]{.65\linewidth}

\begin{tabular}{|l|c|c|c|c|r|}
 \hline
&$k_i$ (m$^{-1})$&$\gamma_i$ (m$^{-1}$ s$^{-1}$) \\
 \hline
$f_1=15$ Hz & 428 &-1.40$\times10^4$
\\
 \hline
$f_2=18$ Hz & 507 &-1.38$\times10^4$
\\
 \hline
$f_3=33$ Hz& 834 &1.24$\times10^4$
\\
 \hline
 \hline
 $f_1=16$ Hz& 455 &-1.65$\times10^5$
\\
 \hline
$f_2=23$ Hz& 626 &-1.58$\times10^4$
\\
 \hline
$f_3=39$ Hz& 946 &1.41$\times10^4$
\\
 \hline
\end{tabular}
\end{minipage}
\caption{(color online) Norms of the wavevectors and interaction coefficients calculated from Eq. \ref{gamma}. The different values are calculated with $\sigma=60$ mN/m and $\rho=1000$\,kg.m$^{-3}$.}
\label{Tab1}
\end{table}
\normalsize
The coefficients of the mother waves are negative whereas the coefficient for the daughter wave is positive. The wave $3$, initially null, would thus grow, by pumping energy from the waves $1$ and $2$. With negligible dissipation, $a_3$ is supposed to grow linearly in time, at short time, as long as $a_3 \ll \sqrt{a_1\,a_2}$. Long time behavior of the system formed by Eqs.~\ref{EqAmp} and \ref{EqPhase} predicts non-linear oscillations and recurrence phenomena~\cite{Simmons1969,Craik}, also with negligible dissipation.\\
\indent
To make a quantitative comparison between these theoretical results and the following experimental measurements, we study the behavior of the daughter wave with the hypothesis $a_3 \ll \sqrt{a_1\,a_2}$. We consider in stationary regime the evolution of the  wave amplitude $a_3$ in the direction $O_\xi$ given by $\mathbf{k_3}$, as previously explained and illustrated in Fig.\ref{triad_graphs1} a). From Eq.~\ref{EqAmp}, we get a linear differential equation of order one with a second member:
\begin{equation}
v_{g3}\, \partial a_3/\partial \xi  = - \delta_3 a_3+ a_{1}a_{2} \, \gamma_3 \sin\phi 
\end{equation}
If the wave-field can be considered homogeneous and if the total phase $\phi$ does not vary with $\xi$, the previous equation can be integrated and by introducing the coordinate $\xi_0$ where $a_3 (\xi_0)=0$, we obtain:
\begin{equation}
a_3 (\xi)=  \dfrac{\gamma_3 \sin\phi}{\delta_3} a_{1}a_{2} \left[1-\exp \left( -\frac{\delta_3}{v_{g3}} \left( \xi-\xi_0\right) \right) \right]
\label{a3}
\end{equation}
To be consistent $ \partial \phi_3/\partial \xi =0$, then Eq.~\ref{EqPhase} imposes $\cos\phi =0$ and thus $\sin \phi = \pm 1$ or $\phi = \pm \frac{\pi}{2}$. The phase locking at $\phi=\pi/2$ was supposed in most of the experimental studies of three-wave resonance~\cite{McGoldrick1970,Henderson1987_1} and was justified by a reasoning of maximal energy transfer. Otherwise the phase $\phi_3$ would be spatially modulated and an analytical solution cannot be expressed any more. The dependency of $a_3$ with the distance $\xi_M$ from the origin, $\xi_0$, is expressed by the prefactor $K$, whose expression is:
 \begin{equation}
K(\xi_M) =1-\exp \left( -\frac{\delta_3}{v_{g3}} \xi_M  \right)
\label{K}
\end{equation}
As a consequence, at a given point $M$, $a_3$ is expected to be proportional to the product $a_1a_2$ and the slope can be identified to $\gamma_3  \sin \phi \, K(\xi_M) / \delta_3$.
In the experiments, $\xi_0$ is assumed to be at the first crossing point between the two mother wave-trains, in the direction of the daughter wave, \textit{i.e.} the origin point $O$ in Fig.~\ref{triad_graphs1} a).
As a remark for surface gravity waves, to validate the four-wave resonant mechanism, experiments were performed by generating two distinct waves trains crossing perpendicularly~\cite{Longuet-Higgins1966,McGoldrick1966}. These experiments with negligible dissipation, validated four-wave resonant interaction in the degenerated case~\cite{Longuet-Higgins1962}, in finite wave-basins. Similarly the beginning of the daughter wave is taken at the first crossing point between the two mother wave-trains.\\
 Homogeneity of the wave-field is hard to fulfil due to viscous dissipation and presence of boundaries, but it will be shown in the following that by taking into account this correction with $K(\xi_M) $, we obtain a satisfying estimation of the interaction parameter $\gamma_3$.
Moreover two limit behaviors can be deduced from Eq.~\ref{a3}. At short distance or for weak dissipation, $a_3$ grows linearly with $\xi$:
\begin{equation}
a_{3\,\mathrm{lin}} (\xi) =\dfrac{\gamma_3 \sin\phi}{v_{g3}} a_{1} a_{2} \left( \xi-\xi_0\right)
\label{a3lin}
\end{equation}
\noindent In contrast at high enough distance, if the amplitude of mother waves remain constant, viscous dissipation can saturate the resonant interaction to a constant value:
\begin{equation}
a_{3\,\mathrm{sat}} (\xi) =\dfrac{\gamma_3 \sin\phi}{\delta_3} a_{1}a_{2} 
\label{a3sat}
\end{equation}
Thus $a_3$ grows with the distance to reach this saturation value due to the balance between non-linear growth and viscous dissipation.
In contrast to the theory which considers an open system, in this work, we investigate three-wave resonance in a closed system, which is the relevant case for capillary wave turbulence experiments. The circular boundaries of the tank reflect indeed a part of incident waves in many directions. The wave-field contains thus a propagative part and a standing (or stationary) part, and this later brings inhomogeneity. In the following, we will show that although effects of reflections are significant, they do not modify the three-wave interaction mechanisms. Thus reflections present in all closed tanks are not preventing resonant interactions, but make their analysis more complex.\\


\section{Experimental set up}

\begin{figure*}
\begin{center}
\includegraphics[width=2.0\columnwidth]{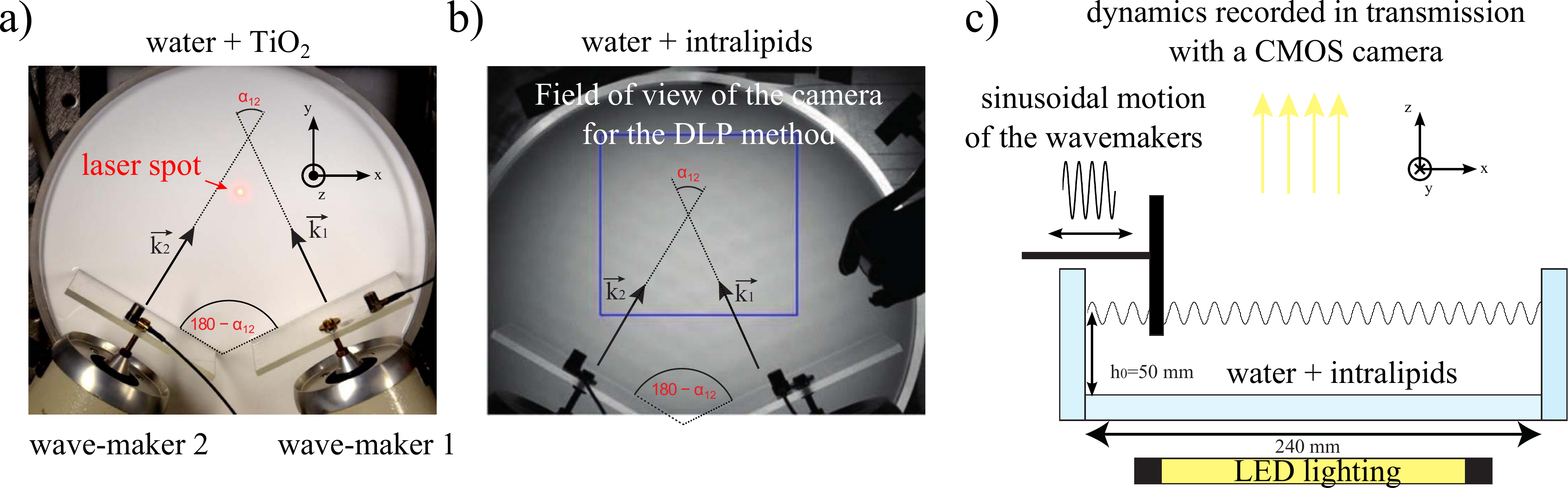}
\caption{(color online) a) Top view of the set up for local temporal measurements: the tank is filled with water mixed with \chemform{TiO_2} pigment. The red spot is the beam of the laser vibrometer. b) Top view of the set up for spatio-temporal measurements with the DLP method. The liquid is a solution of intralipids in water. The blue rectangle depicts the observation area $\mathcal{S}$ of $103 \times 94$\,mm$^2$ where the wave-field reconstruction is performed.  c) Schematic representation of the side view of the set up for DLP spatio-temporal measurements.}
\label{set_up}
\end{center}
\end{figure*}

We use a circular Plexiglas (PMMA) container of diameter $240\,$mm filled with a liquid up to a height $h_0=50$\,mm, corresponding to deep water wave regimes, for frequencies considered in the following. Two gravity-capillary waves $1$ ($\mathbf{k_1},\omega_1$) and  $2$ ($\mathbf{k_2},\omega_2$) are produced at the air-liquid interface using two wave makers, consisting in two vertical rectangular plastic paddles of width $100\,$mm driven horizontally and independently by two electromagnetic shakers (Br\"uel \& Kjaer LDS V201). The paddles are immersed $10\,$mm under the surface of the liquid. Recording of the input excitation is carried out by accelerometers (Br\"uel \& Kjaer 4393) glued on the paddles. Note also, that the distance and the angle between the two wave makers can be tuned. 
\\
Two optical techniques have been used to investigate the wave-field: local measurements with a laser Doppler vibrometer and spatio-temporal measurements using Diffusing Light Photography (DLP) method. The laser Doppler vibrometer (Polytech OPV 506), measures the velocity at one point of the free surface given by the position of the vertical laser beam. The distilled water is white-dyed thanks to \chemform{TiO_2} pigment (Kronos 1001, 10\,g in $1$\,L of water), to make the fluid opaque. The laser beam is thus subjected to a diffuse reflection and the velocity of the point at the free surface is extracted from the interference between incident beam and back-scattered light. After temporal integration, vibrometer can capture deformations smaller than one micron, without any bias. For these concentrations of \chemform{TiO_2}, the properties of pure water at ambient temperature (25$^\circ$C) are weakly modified. We measure a density ratio with water $\rho_{sol\,TiO_2}/\rho_{water}=1.009$ and a kinematic viscosity $\nu=1.02\,10^{-6}\, $m$^2$.s$^{-1}$ using a Anton Paar, MCR 500 rheometer, equipped with a cone-plate combination (diameter 50 mm, angle 1 deg). As \chemform{TiO_2} particles are not accumulating at the interface, surface properties are comparable to those of pure water~\cite{Przadka2012}. By measuring the phase velocity in monochromatic experiments, we find a surface tension of $\sigma=62 \pm 5$\,mN.m$^{-1}$. In the following, measurement point is located at a distance of $90\,\pm 5$\, mm from each wave maker, corresponding to the position $(x_v,\,y_v)$. Figure~\ref{set_up} a) shows a top view of the tank for an angle $\alpha_{12}$ equal to 60 deg, for experiments using the laser Doppler vibrometer placed about $64\,$cm above the liquid surface. As a remark, during preliminary experiments, local measurements were performed using a capacitive wave probe, providing very similar results. Attenuation measurements providing values of the viscous damping coefficients $\delta_i$ are presented in Appendix.\\
 
\indent The second method, Diffusing Light Photography (DLP)~\cite{Putterman1996}, consists in the reconstruction of the 3D free surface from the measurement of the transmitted light through a liquid diffusing the light. Small amount of micrometric particles are added to the liquid ($4.5$\,mL of Intralipids 20\% Fresenius Kabi \texttrademark in $1$L of distilled water). We measure a density ratio with water $\rho_{sol\,intra}/\rho_{water}=1.001$ and a kinematic viscosity $\nu=1.24\,10^{-6}\, $m$^2$.s$^{-1}$. The surface tension will be deduced from the measurements. As displayed in Fig~\ref{set_up} c) a diffuse LED light source of 200$\times$200 mm$^2$ (Phlox) provides a homogeneous lighting below the transparent tank. Transmitted light is recorded on an observation area $\mathcal{S}$ of $103 \times 94$\,mm$^2$, with a fast camera (PCO Edge, scientific CMOS) located above the tank, and with focus made on the surface (frame rate 200 Hz). Knowing that the transmitted intensity is related to the local height of liquid, it is possible with a suitable calibration to reconstruct the wave-field in space and time. More details about DLP are available in an experimental work on capillary~\cite{Berhanu2013}, where a similar set up was used. Spatio-temporal dynamics of free-surface elevation can be thus extracted with a good sensitivity even for steep deformation. This method has an accuracy of order $10\,\mu$m, due to uncertainties in the calibration process, and is thus less precise than the laser Doppler vibrometry. Moreover surfactants present for stabilisation purpose in commercial solutions of intralipids increase slightly wave dissipation (See Appendix). Therefore for quantitative measurements of wave amplitudes, laser Doppler vibrometry will be preferred in the following. But we have checked that the all phenomena observed with the vibrometer are reproduced with the DLP method with less accuracy and slightly different physical constants. DLP will thus succeed to characterize spatial properties of the wave-field and to display mechanisms of three-wave resonant interactions. 

\indent Finally a data acquisition card (NI-USB 6212) controlled through Matlab \texttrademark, sends programmed signals to the electromagnetic shakers and record analog signals from the laser Doppler vibrometer and from the two accelerometers (sampling frequency of $10$\,kHz). As optical properties of liquid are different for the two methods (opaque for the vibrometry and light diffusing for the DLP), local and spatio-temporal measurements are taken separately on dedicated experiments. For the local measurements, recordings are taken during $170\,$s, in which waves are generated for $150\,s$ including transient, stationary and decaying regimes. Each measurement is repeated 12 times to ensure a statistical averaging. Then for spatio-temporal measurements, due to the larger amount of data, the wave reconstruction is performed usually on a duration of $20.5$\,s, in stationary regime.
Throughout the text, the mother wave amplitudes $a_1$ and $a_2$ measured with the vibrometer (wave amplitudes are obtained by bandpass filtering around $f_1$ and $f_2$, see \ref{Amplitudespart}) will be attributed to the spatio-temporal measurement as forcing parameters, which have the same values of forcing amplitude imposed to the electromagnetic shaker. Using this scale, the two kinds of measurements can be compared with a parameter corresponding to properties of the wave-field.

\section{Experimental study}
\label{Classic}
\begin{figure}[h!]
\begin{center}
\includegraphics[width=0.85\columnwidth]{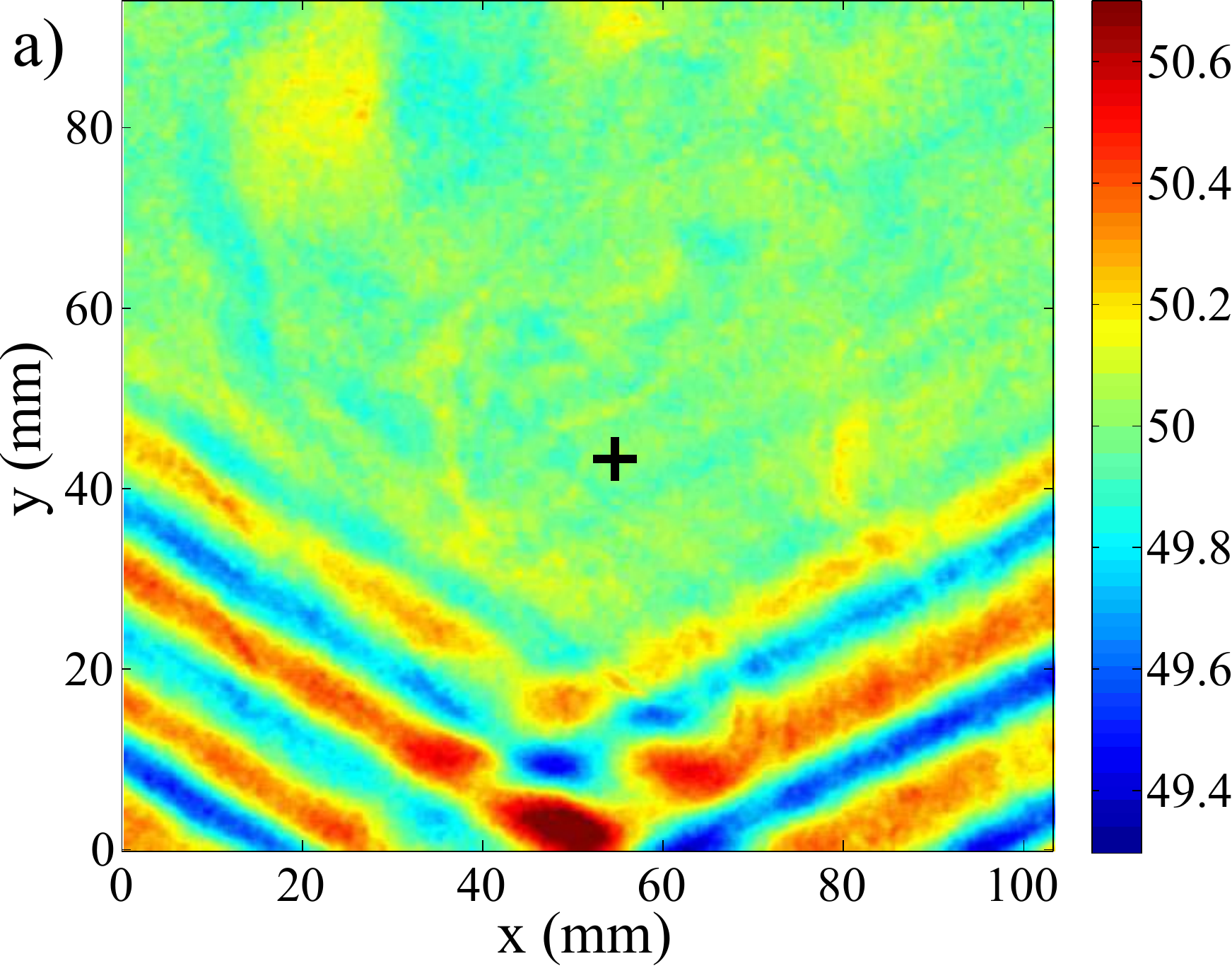}
\hfill
\includegraphics[width=0.85\columnwidth]{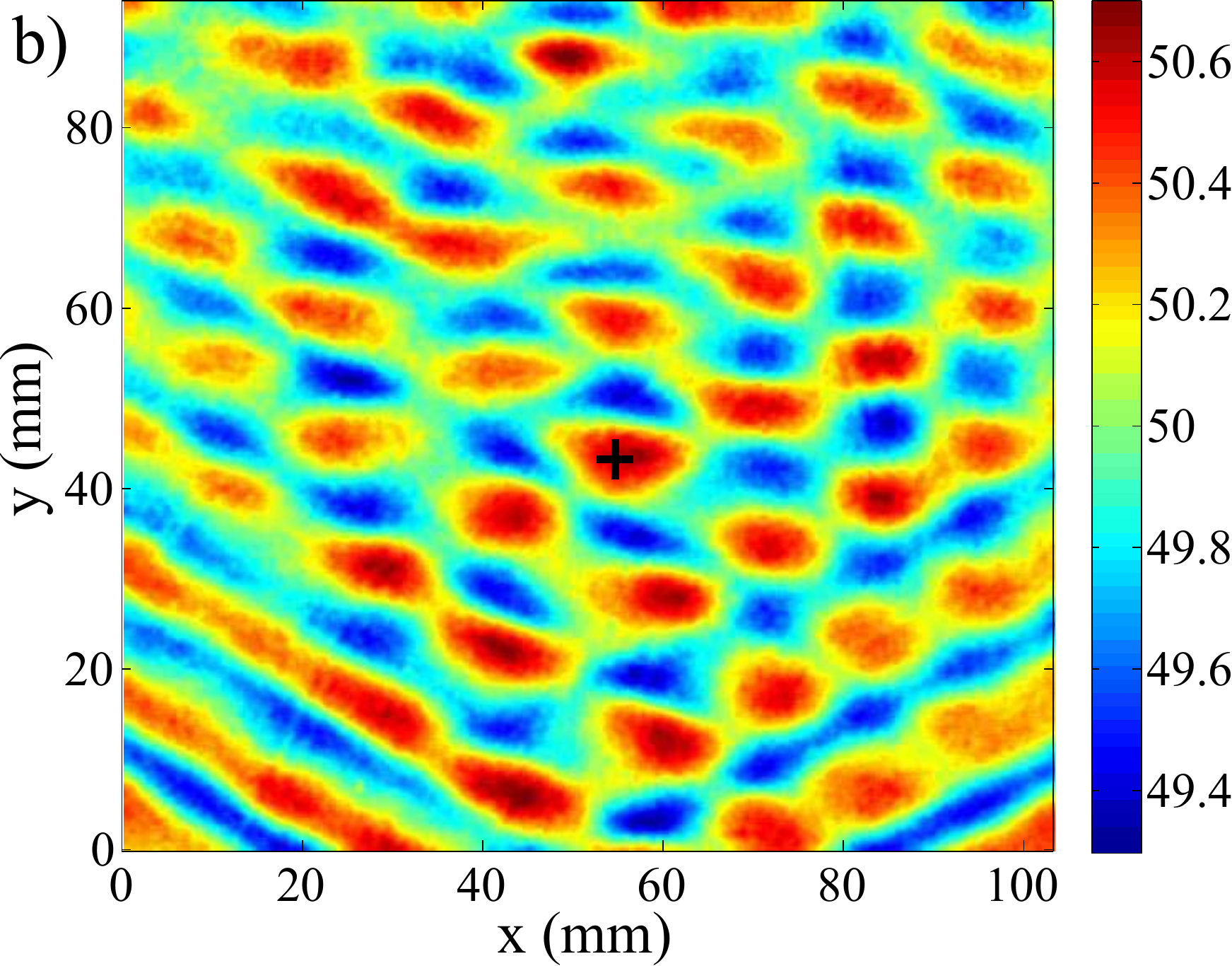}
\caption{(color online) Examples of  wave-field reconstruction obtained with the DLP technique, for $f_1=15$ Hz, $f_2=18$ Hz and $\alpha_{12}=54$ deg: a) in the transient and b) in the stationary regime, with a colorscale in mm (total height of fluid in the tank), with $a_1=130$ and $a_2=132$ $\mu$m (high forcing). The position of the laser spot for vibrometer measurements is indicated by the black cross symbol located at  $x_v=54.6\,$mm et $y_v=43.3\,$ mm. \label{wave-field}}
\end{center}
\end{figure}

We first investigate three-wave interactions, by setting experimentally the conditions imposed by the resonance condition and the dispersion relation. Here we report the results for $f_1=15$ and $f_2=18$ Hz, imposing the angle $\alpha_{12\,r}=54$ deg from Eq.~\ref{Eqalpha}, between the two wave-trains. Appearance of a third wave at the frequency $f_3=f_1+f_2=33\,$Hz is thus expected. The two wave-trains are generated by the sinusoidal motion of each paddle. The angle between the two paddles is set at  $180-54=126$ deg with an experimental accuracy estimated to be $\pm 2$ deg. Examples of the wave-field obtained with DLP measurements, in the transient and the stationary regimes, are displayed in Fig.~\ref{wave-field}. In the interaction zone between the two mother waves, a modulated wave-field is observed, with crests and troughs in the surface height distribution, corresponding mainly to the linear superposition of the two mother waves. To detect the presence of the daughter wave, performing a Fourier transform analysis is necessary. From DLP measurements, spatial homogeneity can be also estimated by computing the ratio between the spatial standard deviation of wave amplitude and the spatial average wave amplitude and typical values of $10\%$ are found. The control parameters will be the amplitudes of mother waves $a_1$ and $a_2$.

\subsection{Power spectra of wave elevation}
\begin{figure}[h!]
\begin{center}
\includegraphics[width=0.85\columnwidth]{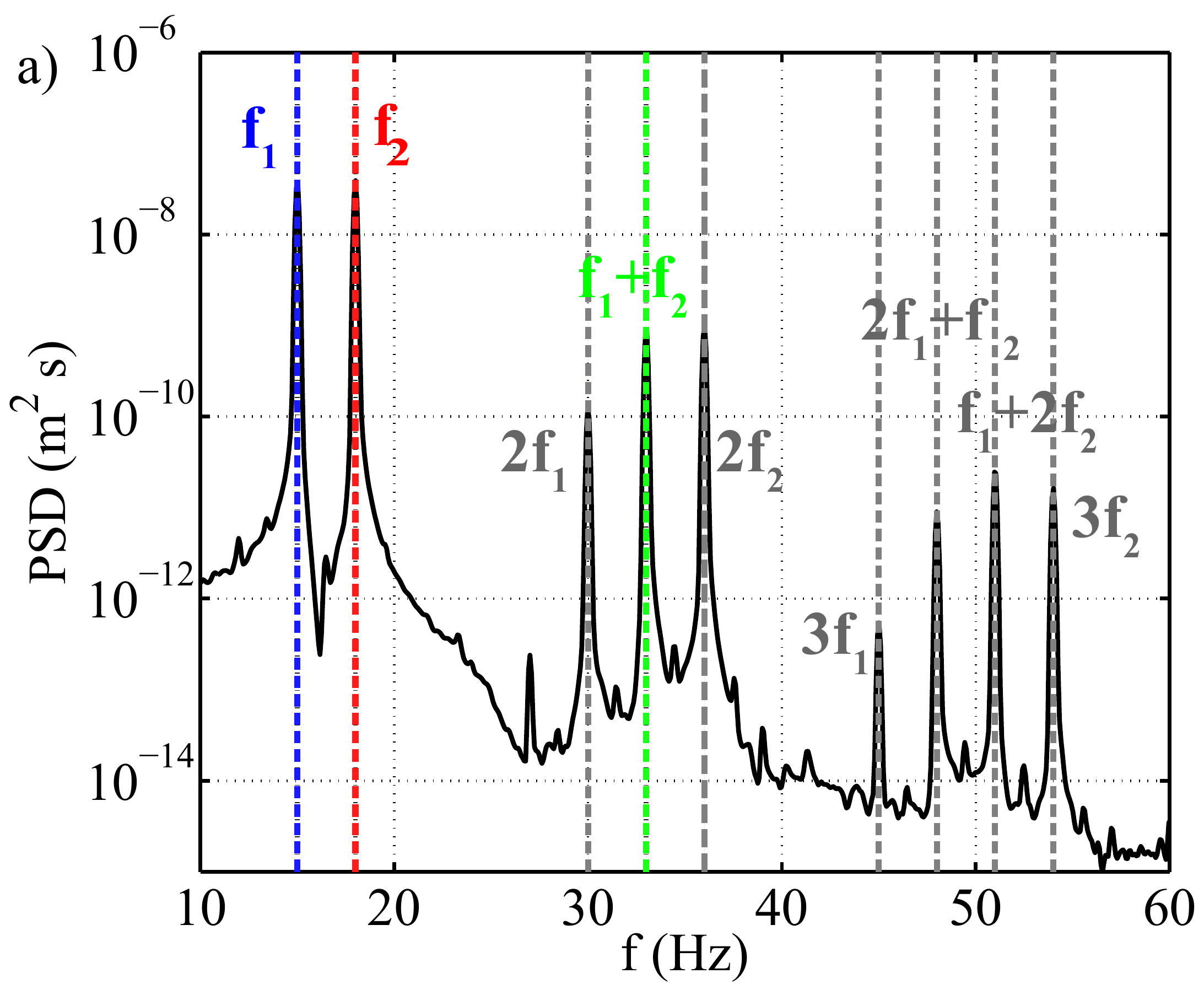}\hfill
\includegraphics[width=0.85\columnwidth]{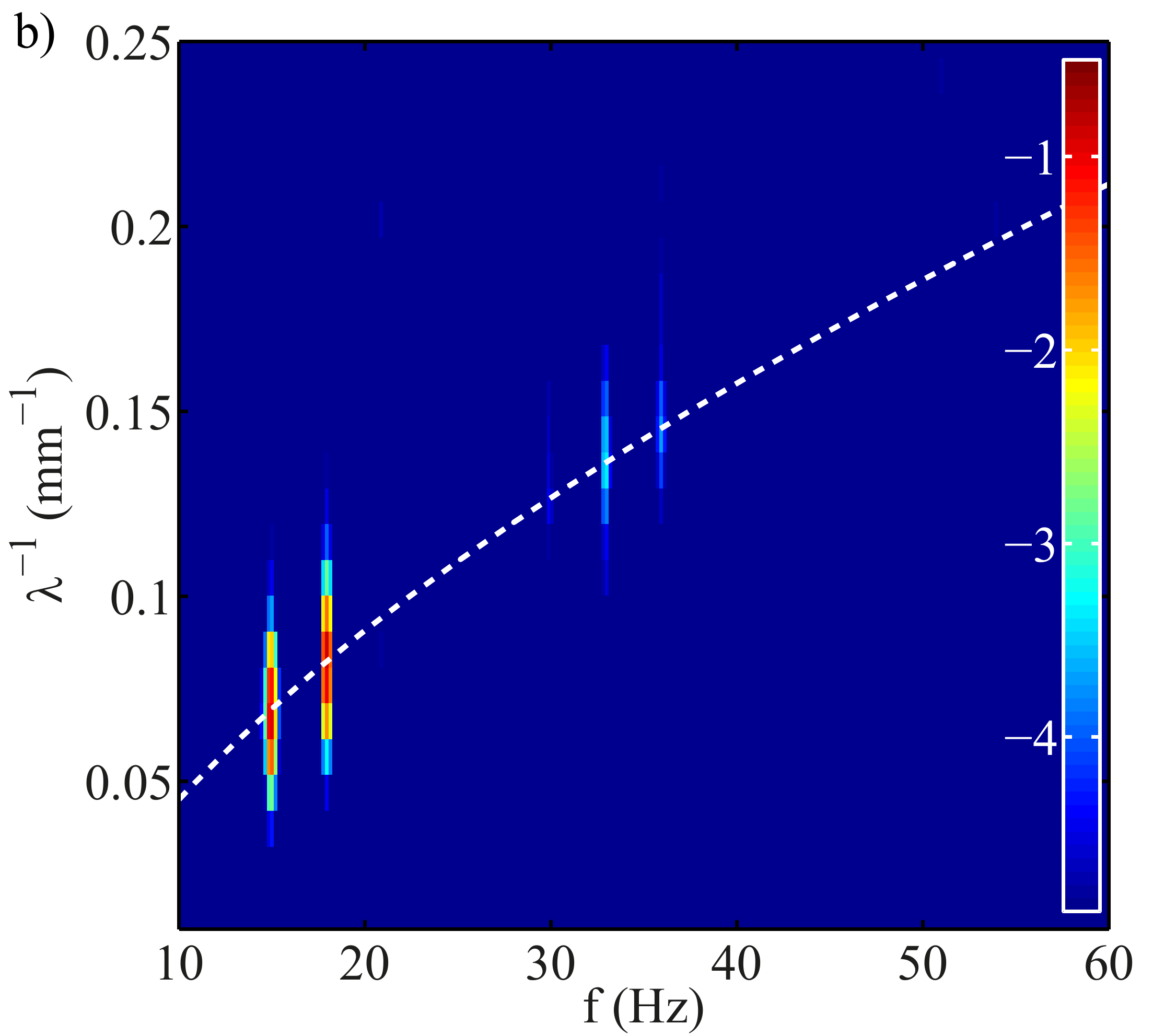} 
\caption{(color online) a) Local Power Spectrum $P_\eta(\omega)$ for $f_1=15$, $f_2=18$ Hz and $\alpha_{12}=54$ deg. There, $a_1=130$ and $a_2=132$ $\mu$m (high forcing). b) Spatio-temporal spectrum of wave elevation $S_\eta (\omega,k)$.  White dashed line: linear theoretical dispersion relation for $\sigma_{fit}=55\,$mN/m. Colorscale corresponds to $\log_{10}(S_\eta (\omega,k))$ with $S_\eta$ in mm$^3.$s, with $a_1=55$ and $a_2=56$ $\mu$m (low forcing).}\label{spectra}
\end{center}
\end{figure}

We detect the presence of the daughter wave by computing the local power spectrum $P_\eta (\omega)$ of wave height $\eta(t,x_v,y_v)$, using the vibrometer. $P_\eta$ is defined as the square modulus of the Fourier transform of $\eta$ and is computed using the \textit{pwelch} function with Matlab \texttrademark. A typical spectrum for an intermediary amplitudes among the tested one is shown in Fig.~\ref{spectra} a). Peaks corresponding to the mother waves are clearly visible at frequency $f_1$ and $f_2$. An additional peak of smaller amplitude is present at the frequency $f_3=f_1+f_2$,  clearly demonstrating the existence of a mode at $f_3$. Note here that the amplitude of the daughter wave at $f_3$ is of the same order of magnitude as the amplitudes of the two harmonics $2f_1$ and $2f_2$. These harmonics are generated by non-linearity in the generation by the wave makers (typical mother wave steepnesses: approximately ($0.01 < k_i \, a_i < 0.1$). A peak of smaller amplitude (not depicted) is also observed at $f=f_2-f_1=3\,$Hz. Although its generation is also related to a three-wave mechanism, the corresponding wave-length (around $167$\,mm) is too close to the tank dimensions and we choose to focus on the triad, which is transferring energy at high frequency. Moreover especially when the forcing amplitude is high enough, higher order resonant triads are also observable in the spectrum, like: $(f_1,2f_1,3f_1)$, $(f_1,f_1+f_2,2f_1+f_2)$, $(f_2,2f_1,2f_1+f_2)$,$(f_2,f_1+f_2,f_1+2f_2)$, $(f_1,2f_2,f_1+2f_2)$ and $(f_2,2f_2,3f_2)$, which can be seen in Fig.~\ref{spectra} a). At high forcing amplitude, spectral peaks become wider, due to non-linear broadening. Therefore at high enough forcing the considered triad cannot be taken isolated and we expect a departure from the theoretical considerations of Sect. \ref{Amplitudes}.

Using DLP method, we compute also spatio-temporal power spectra $S_\eta(\omega,k_x,k_y)$ by performing a temporal 2D spatial Fourier transform and a temporal Fourier transform on the wave-field $\eta(x,y,t)$. The spectrum $S_\eta (\omega,k)$ with $\, k=\sqrt{{k_x}^2+{k_y}^2}$ is averaged over the different directions and is displayed  in Fig.~\ref{spectra} b), as a function of the inverse of the wavelength $\lambda$ and of the frequency $f$. The spatial resolution of the spectrum is equal to $2 \pi \,\delta \lambda^{-1}= \delta {k} \approx 30.5\,$m$^{-1}$. Peaks of the spectrum appear as high amplitude spots. As seen previously, spectrum contains peaks at the frequencies of the mother waves $f_1$ and $f_2$, the harmonics of the mother waves $2f_1$ and $2f_2$ (hardly visible) and the daughter wave $f_3=f_1+f_2$. 
Small amplitude waves are expected to follow the linear gravity-capillary dispersion relation. The experimental dispersion relation can thus be accurately fitted by the linear dispersion relation where the surface tension $\sigma$ is the only free parameter in Eq.~\ref{disp}. We find here $\sigma_{fit}=55\,$mN/m, but for other experiments with solution of intralipids $\sigma$ can be significantly lower, up to $45$\,mN/m. We observe also a broadening of the relation dispersion in ${\lambda}^{-1}$, evaluated using a gaussian adjustment of the peaks, as $2 \pi \,\delta \tilde{\lambda}^{-1}= \delta \tilde{k} \approx 70\,$m$^{-1}$  around the fit. This broadening is due to the finite field of view of the camera, to the weak non-linearity and to the dissipation of the waves. 

\subsection{Verification of the spatial resonance condition} 

 \begin{figure*}
 \begin{center}
\includegraphics[width=0.66\columnwidth]{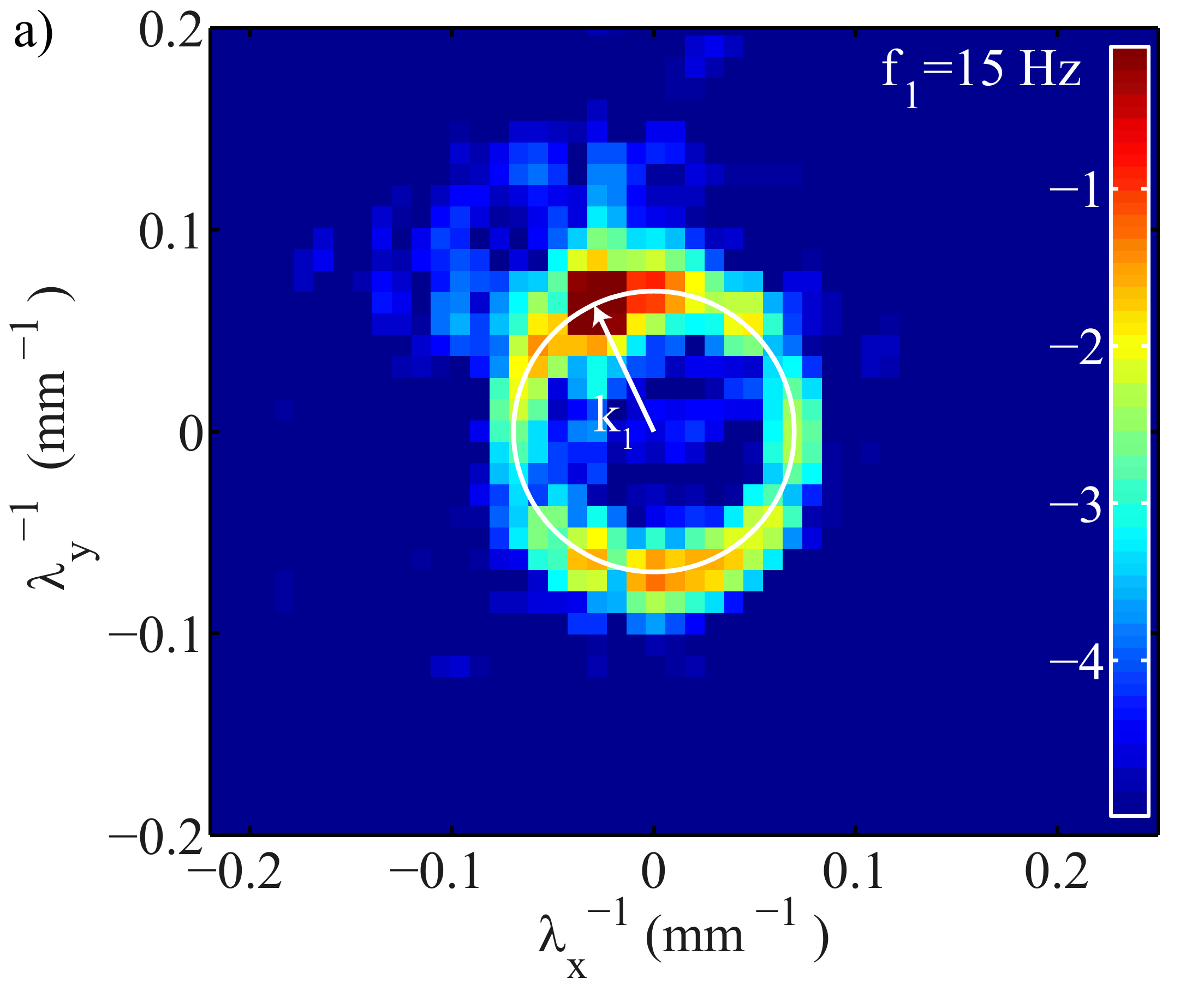}
\includegraphics[width=0.66\columnwidth]{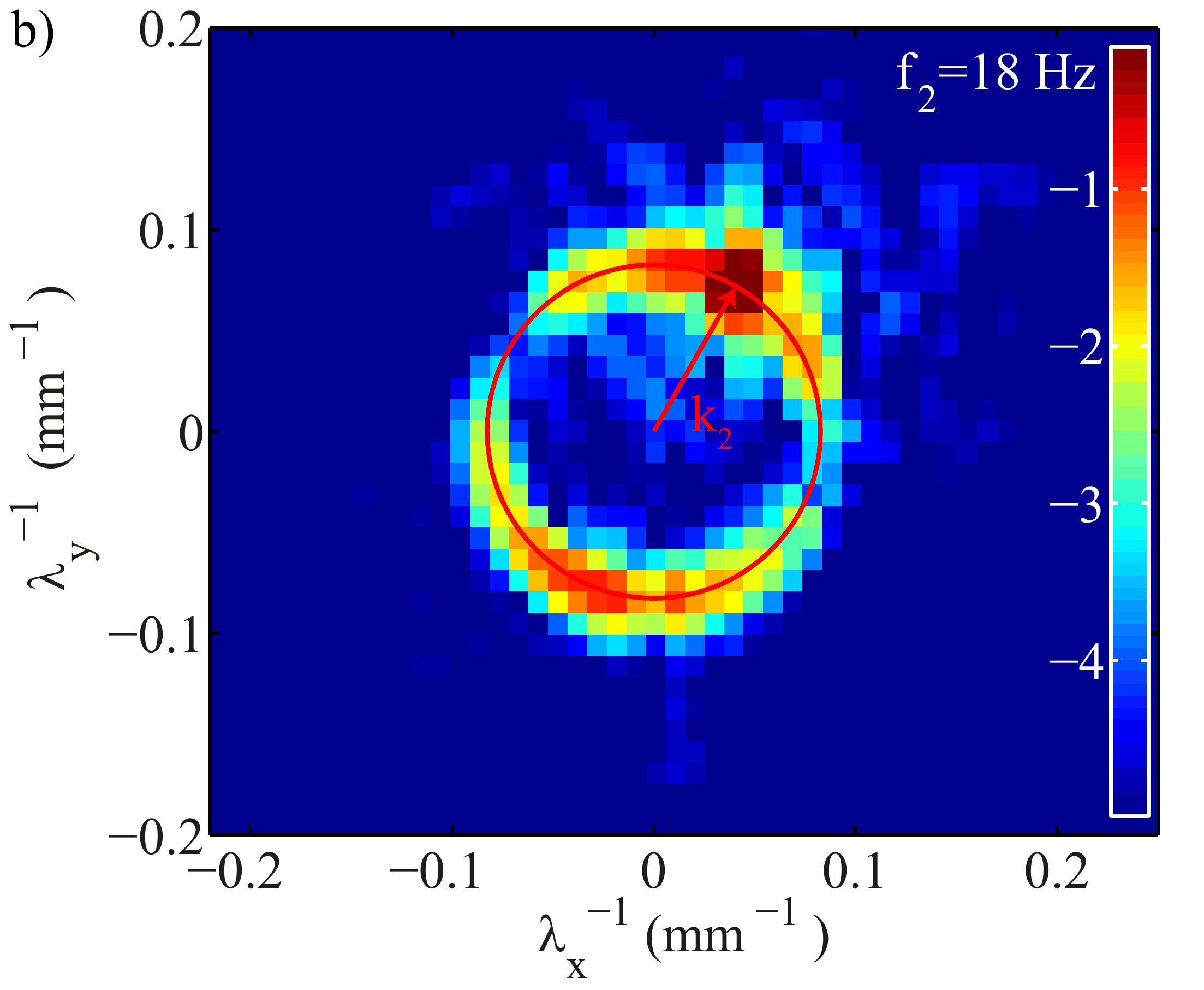} 
\includegraphics[width=0.66\columnwidth]{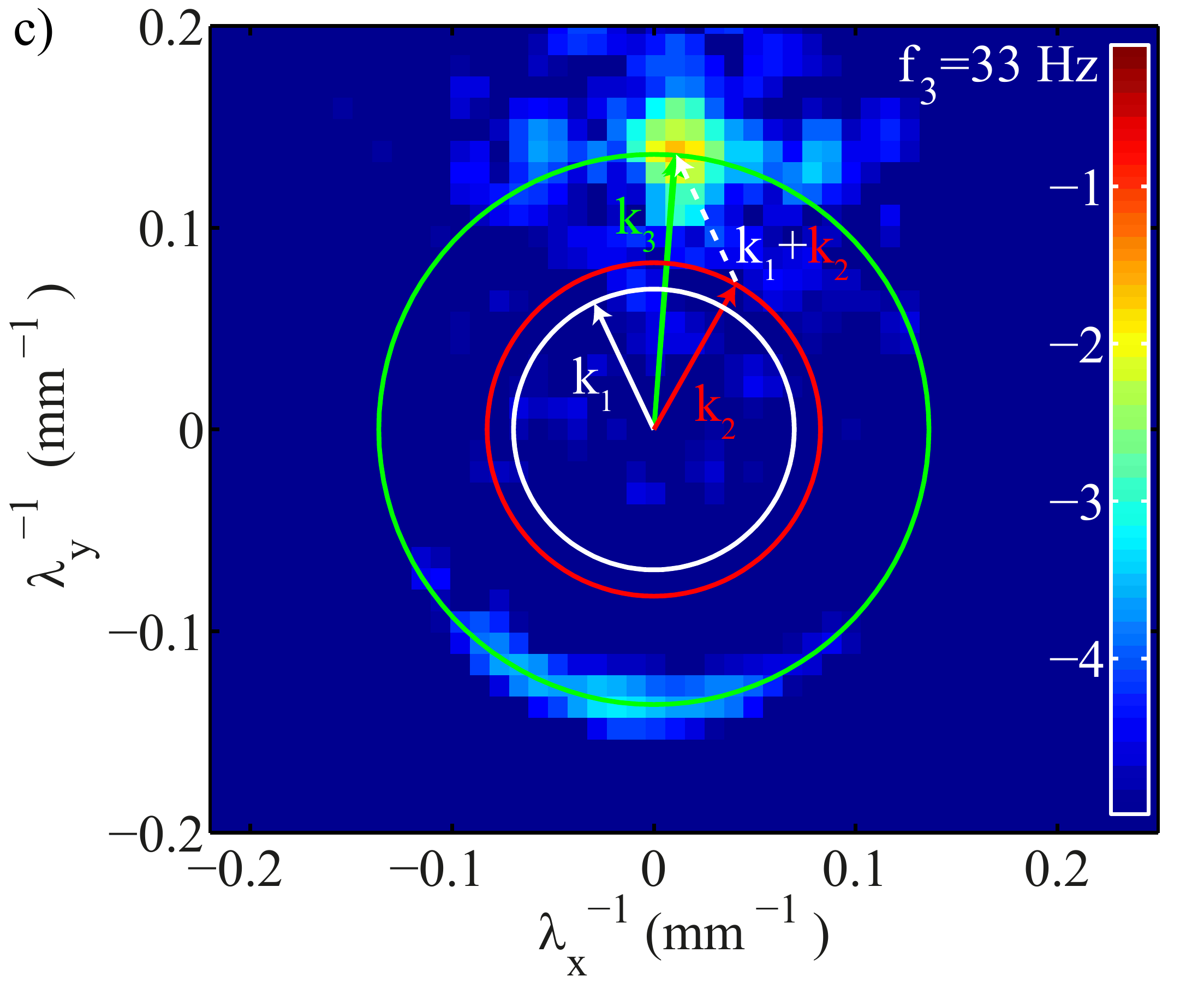}
                \caption{(color online) 
Spatio-temporal spectra of wave elevation $S_\eta (\omega,k_x,k_y)$,  is displayed for the triad frequencies: a)  $f_1$,  b) $f_2$ and c) $f_3$. Circles correspond to the linear theoretical dispersion relation at the given frequency. The arrows indicate the vector $\mathbf{k_i}$ extracted from the maxima of the experimental spectra. We observe $\mathbf{k_3}\approx \mathbf{k_1}+\mathbf{k_2}$.  Colorscale $\log_{10}(S_\eta)$ arbitrary unit, with $a_1=55$ and $a_2=56$ $\mu$m.}
    \label{sp_spectra_1518_54d_all}
       \end{center}
 \end{figure*}  
  \begin{figure}[!h]
 \begin{center}
 \includegraphics[width=.85\columnwidth]{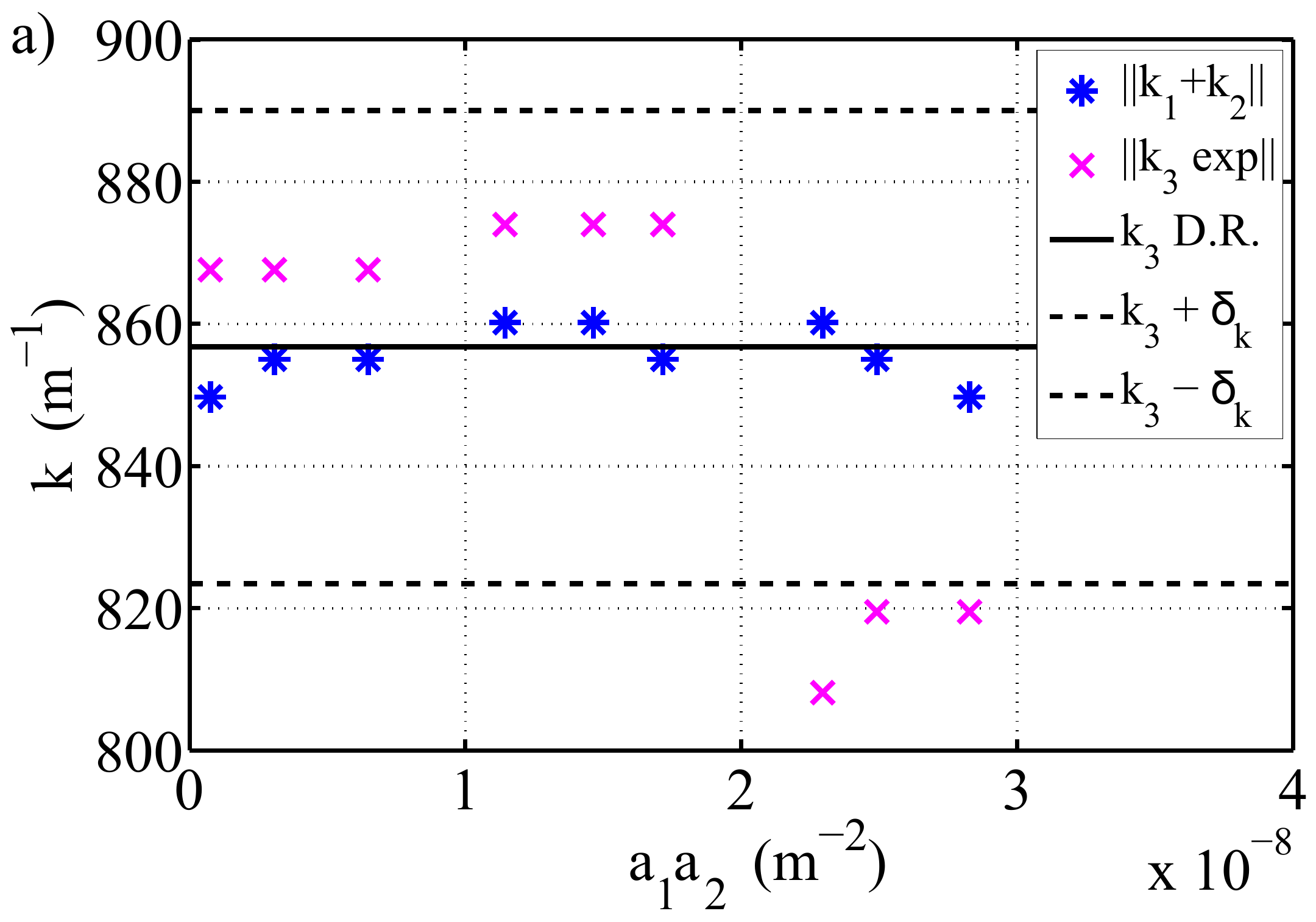}  \hfill

 \includegraphics[width=.85\columnwidth]{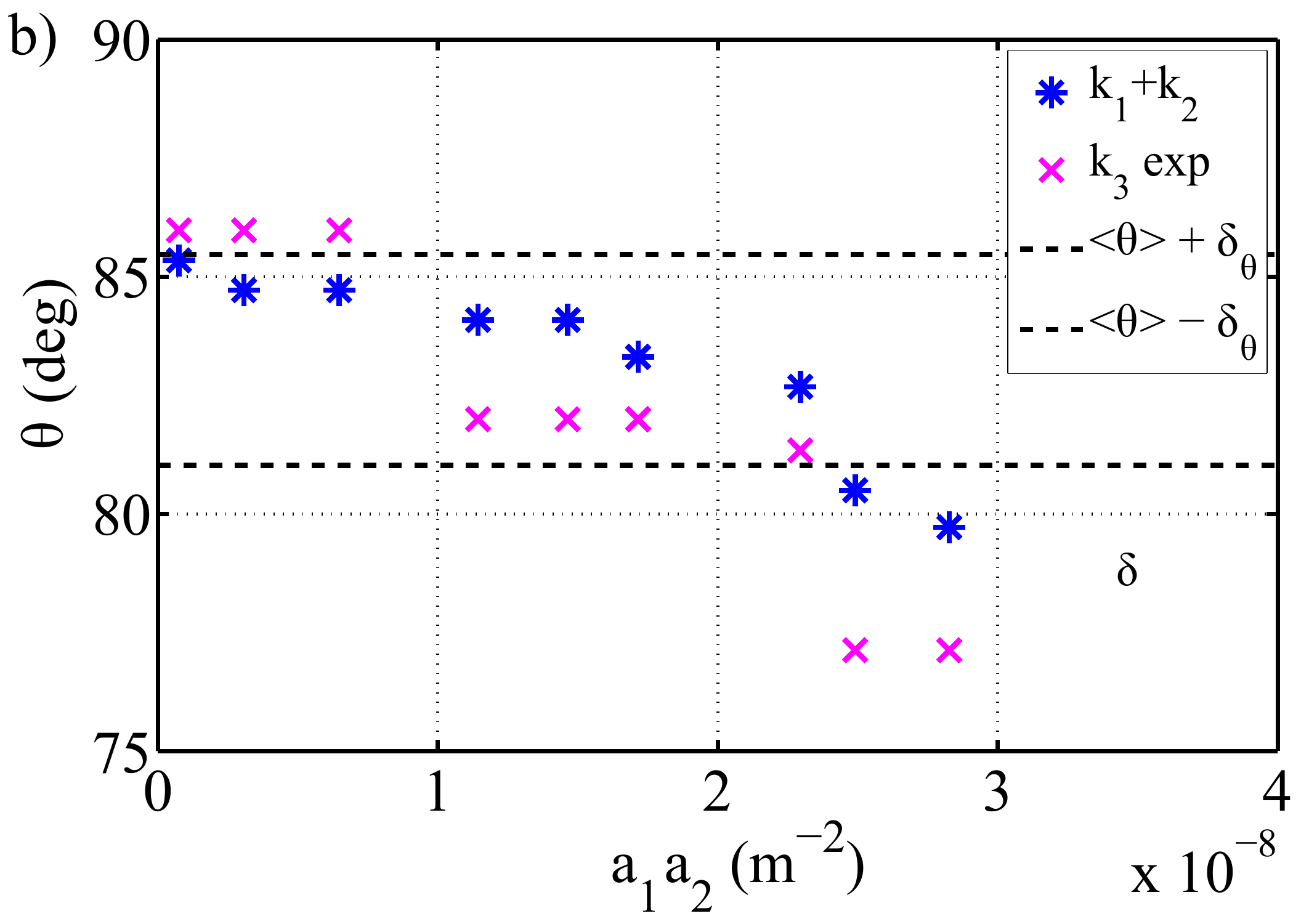} 
        \caption{(color online) Comparison between the experimental values of $\mathbf{k_1}+\mathbf{k_2}$ and $\mathbf{k_3\,\mathrm{exp.}}$ as a function of the product $a_1\,a_2$ of the mother wave amplitudes. a) in norm and b) in direction, when $\alpha_{12}=54$ deg. In a) the value of $k_3$ given by the linear dispersion relation is represented by a black line. The black dashed line provides the acceptable bounds due to the finite resolution $\delta_k= 30.5$ m$^{-1}$ of the spectra. In  b) $\theta$ is the angle between the horizontal axis of Fig.~\ref{sp_spectra_1518_54d_all} c) and $\mathbf{k_3}$. The black dashed lines indicate the angular accuracy $\delta_\theta$ around the average orientation $\left\langle \theta \right\rangle $ of $\mathbf{k_1}+\mathbf{k_2}$. The product $a_1a_2$ has been measured with the laser vibrometer, in equivalent experimental conditions. }
            \label{accordtriad1518_54}
        \end{center}
 \end{figure}    
Resonance implies that both temporal and spatial conditions $f_3=f_1+f_2$ and $\mathbf{k_3}=\mathbf{k_1}+\mathbf{k_2}$ are satisfied simultaneously. Fig.~\ref{sp_spectra_1518_54d_all} shows separately the spatio-temporal spectra $S_\eta(\omega_i,k_x,k_y)$ for the frequency $f_1$, $f_2$ and $f_3$. In Fig.~\ref{sp_spectra_1518_54d_all} a) and b), maximum of each spectrum is observed in the direction of propagation of the wave 1 or 2 at a position from the center equal to $k_i=\parallel \mathbf{k_i} \parallel$ given by the linear dispersion relation. But wave energy is also detected with wavenumber $k_1$ and $k_2$ in other directions than the initial propagation one. Since measurements are performed during the stationary regime, this is due to the multiple reflections on the border of the tank. Despite the multidirectionality observed for the mother waves, for the daughter wave at $f_3$ in Fig.~\ref{sp_spectra_1518_54d_all} c), a maximum is clearly detected at $k=k_3$ and the corresponding wavevector $\mathbf{k_3}$ is close to the vectorial sum of $\mathbf{k_1}$ and $\mathbf{k_2}$.\\

\begin{figure*}
\begin{center}
\includegraphics[width=0.66\columnwidth]{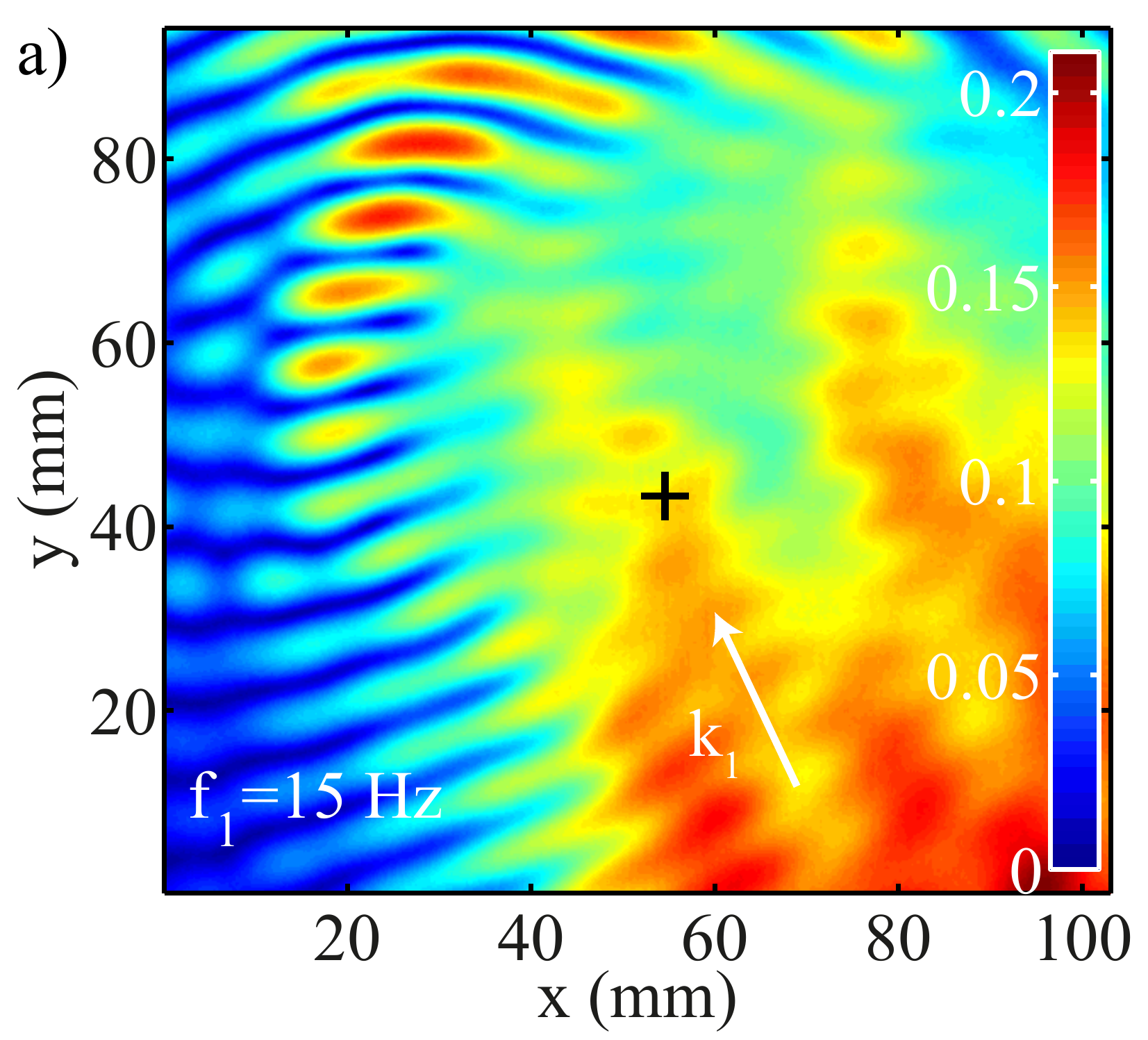} \hfill
\includegraphics[width=0.66\columnwidth]{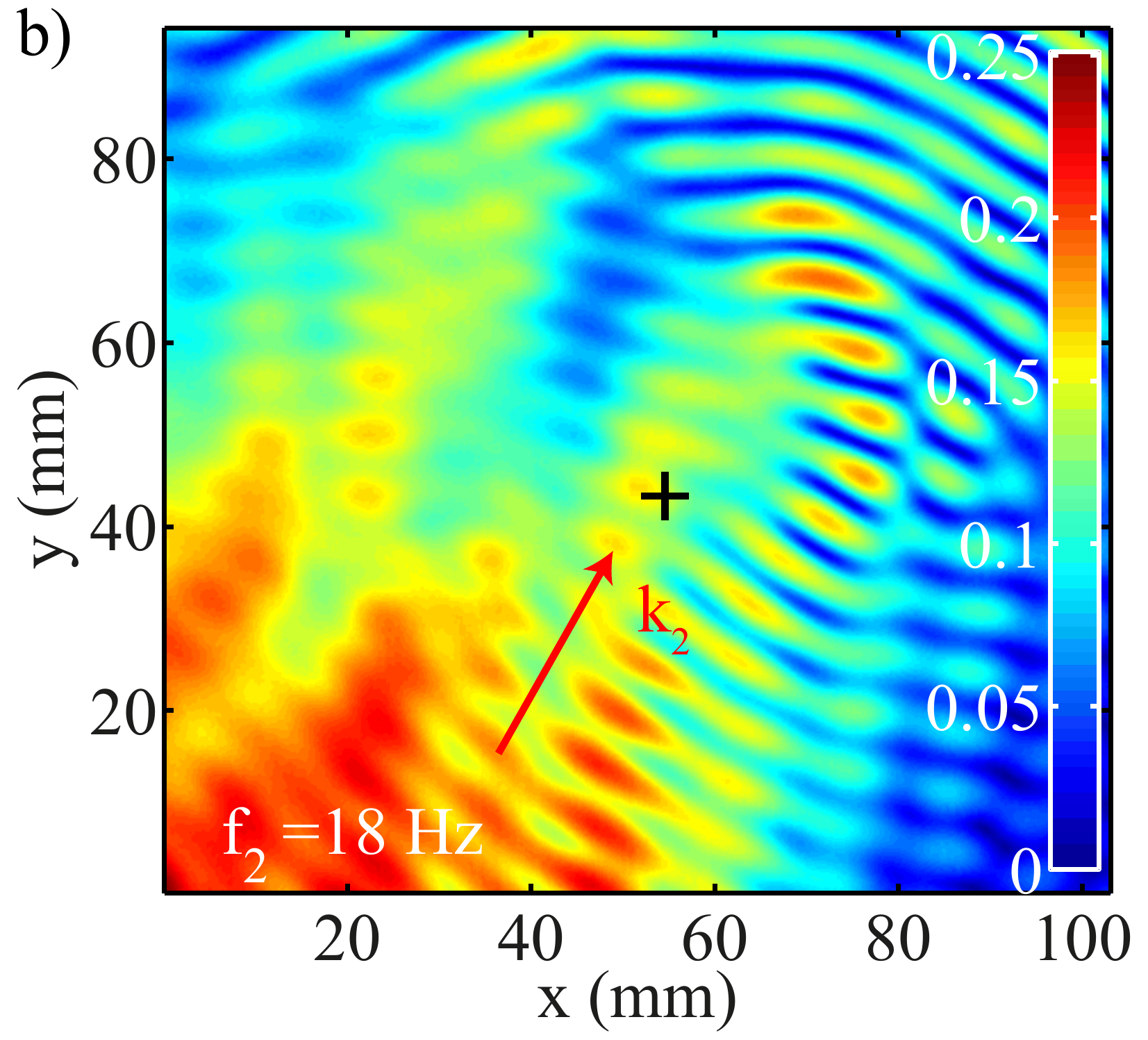} \hfill
\includegraphics[width=0.66\columnwidth]{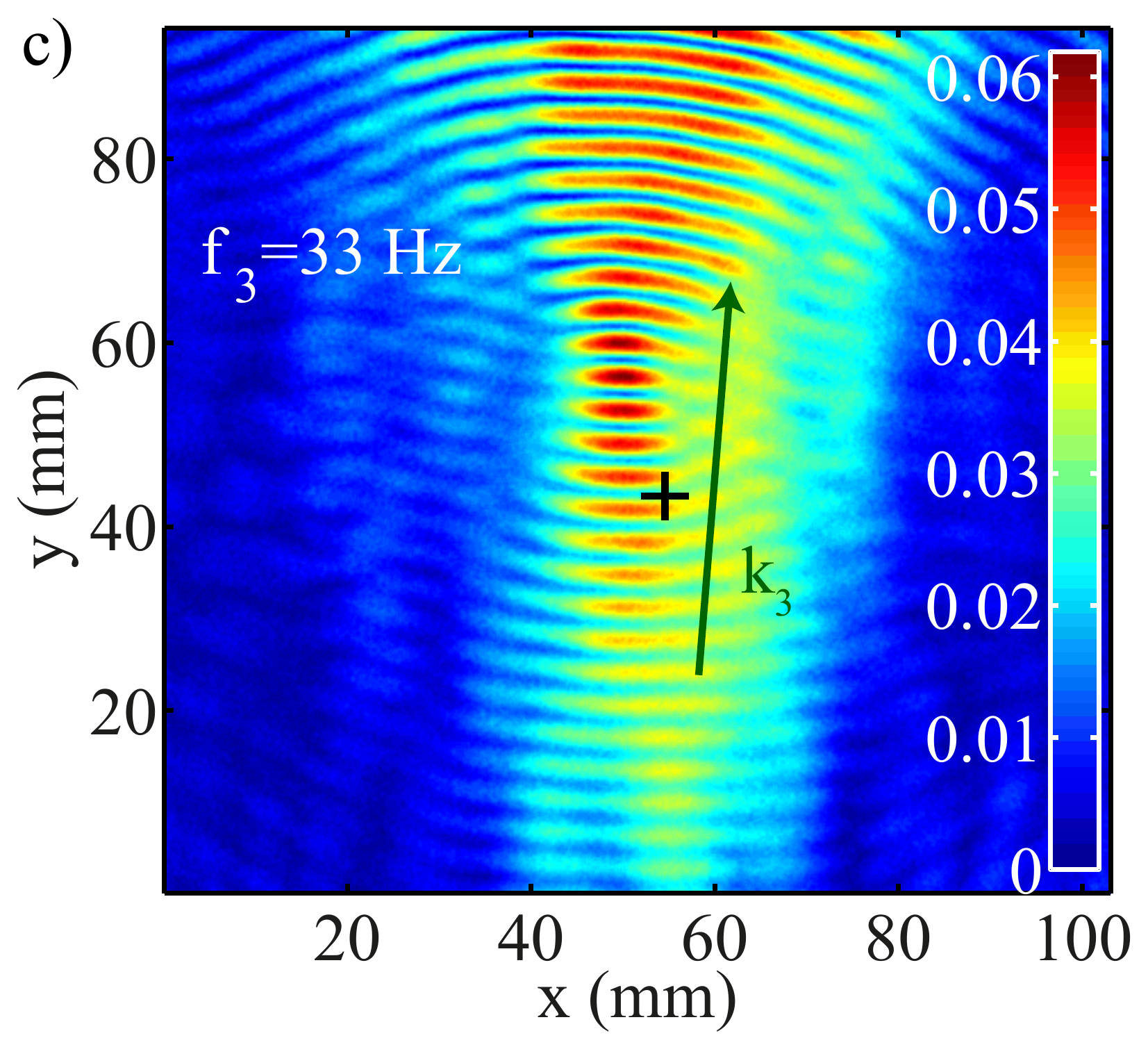} 
\caption{a) $\tilde{a}_1(x,y)$ providing the spatial distribution of wave-mode at the frequency $f_1$, when $\alpha_{12}=54$ deg. The black cross indicates the corresponding position of the laser beam for the measurements performed with the vibrometer. Colorscale $\tilde{a}_i$ in mm. Here $a_1=104$ and $a_2=110$ $\mu$m.  b) $\tilde{a}_2(x,y)$ spatial mode at $f_2$.  c) $\tilde{a}_3(x,y)$ spatial mode at $f_3$. }
\label{spatialmodes1}
\end{center}
\end{figure*}
\noindent

\begin{figure}[h!]
\begin{center}
\includegraphics[width=0.85\columnwidth]{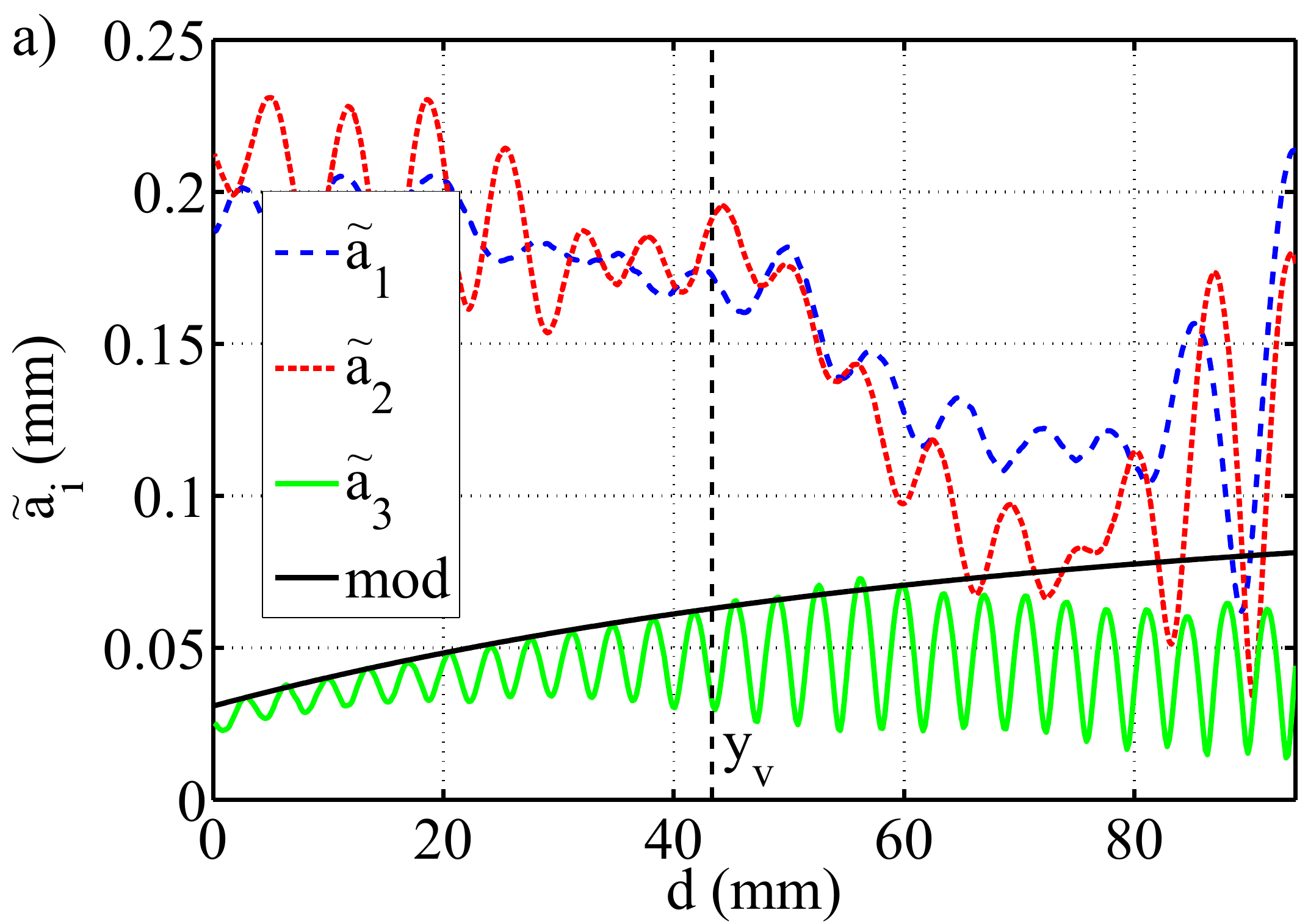} 
\hfill
\includegraphics[width=0.85\columnwidth]{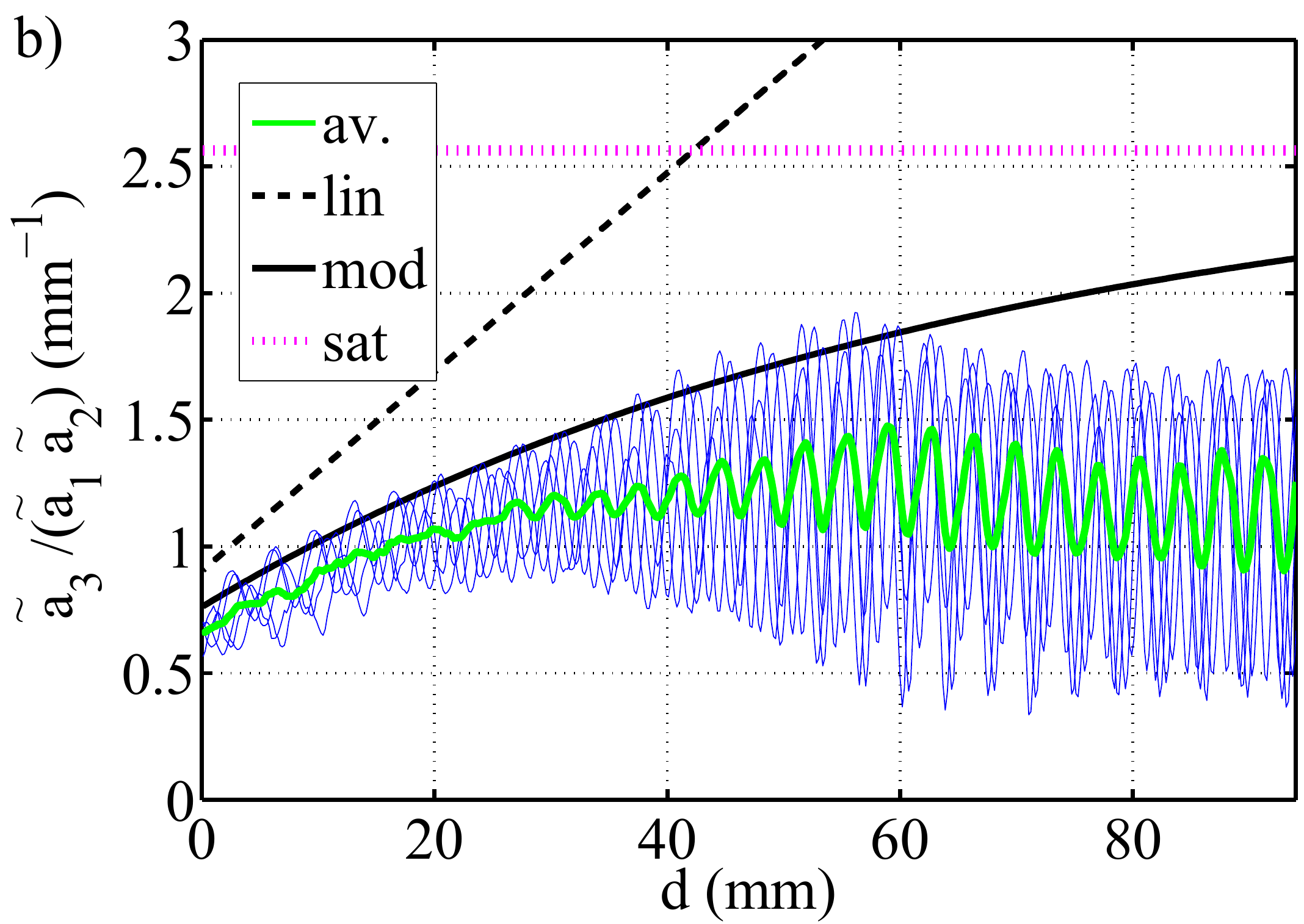} 
\caption{(color online) a) Evolution of wave amplitudes $\tilde{a}_i$ (averaged between $49<x<55$\,mm)\,along the propagation direction of the wave $3$ as a function of the distance $d$ from the bottom of the image. The behavior of $\tilde{a}_3$ is compared to the model (black solid line)  given by Eq.~\ref{a3} with $\xi \simeq d$. The vertical black dashed line gives the position $y_v$ of the laser beam for vibrometer measurements. b) Rescaled amplitude of $\tilde{a}_3 /(\tilde{a}_1 \tilde{a}_2)$ as a function of the distance $d$ for different measurements ($6.5\, 10^{-9} < a_1 a_2 < 1.7\, 10^{-8}\,$m$^2$). We notice that the spatial amplitude modulations are not in phase in the different measurements. The average evolution of $\tilde{a}_3$ on the measurements, green (light gray) line, is compared with the model given by Eq.~\ref{a3}, black solid line. The linear limit for negligible dissipation (black dashed line from Eq.~\ref{a3lin}) and the saturated value (light-gray dotted line from Eq.~\ref{a3sat}) are also depicted. }\label{spatialmodes2}
\end{center}
\end{figure}
\noindent
To quantify the verification of the spatial resonance condition, a systematic study was performed for growing mother wave amplitudes. In Fig.~\ref{accordtriad1518_54} for increasing $a_1a_2$, we test the accordance of $\mathbf{k_3}$ with the  dispersion relation and on the direction $\theta$ made by $\mathbf{k_3}$ with the horizontal axis.
As the spectra have a finite resolution $\delta_k = 30.5$\,m$^{-1}$, the modulus of $\mathbf{k}$ is measured with an accuracy equal to $\pm \delta_k$ and its direction with an accuracy equal to  $\pm \delta_\theta =\delta_k /k_3 = 2.0$ deg. The resonant condition is thus well verified for low values of the product $a_1 a_2$. When this product becomes larger than $2\, 10^{-8}$\,m$^2$ a significant departure is observed, probably due to higher non-linear effects. The angle $\theta$ is also decreasing with the amplitude of the two mother waves. Nevertheless spatio-temporal measurements show that resonance conditions are well verified in time and space for a significant range of mother-wave amplitudes.

\subsection{Spatial behavior of mother and daughter waves}

Using the spatio-temporal DLP measurements, we aim to access to the spatial behavior of each component in the triad. Indeed according to Eq.~\ref{a3}, the daughter wave is expected to grow with the distance and thus its amplitude depends on the spatial coordinate. Moreover as the system is finite and presents inhomogeneity due to reflections and viscous dissipation, the spatial evolution of the waves has to be studied. To do this, Fourier power spectrum is applied only in time to the wave-field $\eta(x,y,t)$, to obtain a spectrum $S_\eta (x,y,f)$, which is here a function of spatial coordinates and frequency. By taking the spectra for the frequencies $f_i$ in the triad, we can define wave-modes $\tilde{a}_i(x,y)$ depicting the spatial distribution of each wave of frequency $f_i$ in the triad. To express $\tilde{a}_i$ as an amplitude, they are obtained by integration of the spectrum around $f_i$, $\tilde{a}_i(x,y)=\sqrt{2}\,\left(\int^{f_i+\delta_f}_{f_i-\delta_f} S_{\eta}(x,y,f_i)\,df \right)^{1/2}$\,with $i=1,2,3$ and $\delta_f=0.2$\,Hz. Note by definition, that $\tilde{a}_i$ is averaged with time. Spatial wave-modes $\tilde{a_1}\,$, $\tilde{a}_2$ and $\tilde{a}_3(x,y)$ are plotted respectively in Fig.~\ref{spatialmodes1} a), b) and c). Due to the circular boundary of the tank producing a stationary component on the wave-field, on each graph, we observe a significant modulation at the wavelength $\lambda_i/2$ for the frequency $f_i$, because the power spectrum is a quadratic operation. Nevertheless $\tilde{a}_i$ provide a useful evaluation of local wave amplitudes. We observe that the inhomogeneity of the mother wave modes are significant. Nonetheless in agreement with the results from the spatial spectra, the wave $3$ is found to propagate in a direction close to $O_y$ and its amplitude is important in the middle of the crossing region between the two mother wave-trains. Then by plotting the amplitudes $\tilde{a_i}$ as a function of the distance $d$ from the bottom of the image, Fig.~\ref{spatialmodes2} a), we observe a decay of $\tilde{a}_1$ and $\tilde{a}_2$ due to viscous dissipation and non-linear pumping by the wave $3$. $\tilde{a}_3$ is indeed found to grow slightly. This behavior is qualitatively described with Eq.~\ref{a3}, by taking $\xi \simeq d$. In this equation $a_1 a_2$ is taken as the average value of $\tilde{a}_1 \tilde{a}_2$ for $d<30$\,mm. We assume that $\sin \phi =1$ and we take $v_{g3}=0.316\,$m.s$^{-1}$, $\delta_3=4.83$\,s$^{-1}$ and so that $\gamma_3$ is taken equal to $1.30 \,10^4\,$m$^{-1}$ (see Eq.~\ref{gamma} with the physical properties of the Intralipid solution). Then the  origin $O$ at $\xi=0$ is determined by matching $\tilde{a_3}$ for $d=0$, which gives $d_0=-23\,$mm. Although $\tilde{a}_3$ is not so small for this measurement and that the $\tilde{a}_1$ and $\tilde{a}_2$ are varying in space, the model depicts roughly the $a_3$ evolution with the good order of magnitude for $d < 50$\,mm. At higher distance, inhomogeneity seems too important to make a quantitative comparison.  

These results are confirmed for several measurements with different forcing amplitudes by plotting the rescaled amplitude $\tilde{a}_3 /(\tilde{a}_1 \tilde{a}_2)$ in Fig.~\ref{spatialmodes2} b). We observe that the fast spatial modulations are not coherent from a measurement to another, although experiments have been performed successively in identical conditions. We suppose that the stationary pattern of standing waves depends in the capillary regime on the meniscus shape on the border of the circular tank, which is known to be subjected to hysteresis. Except at lowest mother wave amplitudes, where signal to noise ratio is too important and at high amplitude, the rescaled evolution of $\tilde{a}_3$ is in qualitative agreement with the model despite the approximations, which validates the resonant interaction theory. Moreover we demonstrate that for capillary-gravity waves, neither the linear growth solution (Eq.~\ref{a3lin}) nor the saturated solution by viscous dissipation (Eq.~\ref{a3sat}) describes the observations and thus the more complete solution (Eq.~\ref{a3}) has to be used, which is taking into account the spatial growth of $a_3$ with $\xi$.

\subsection{Amplitudes and phase locking in stationary regime}
\label{Amplitudespart}

\begin{figure*}
\begin{center}
\includegraphics[width=.85\columnwidth]{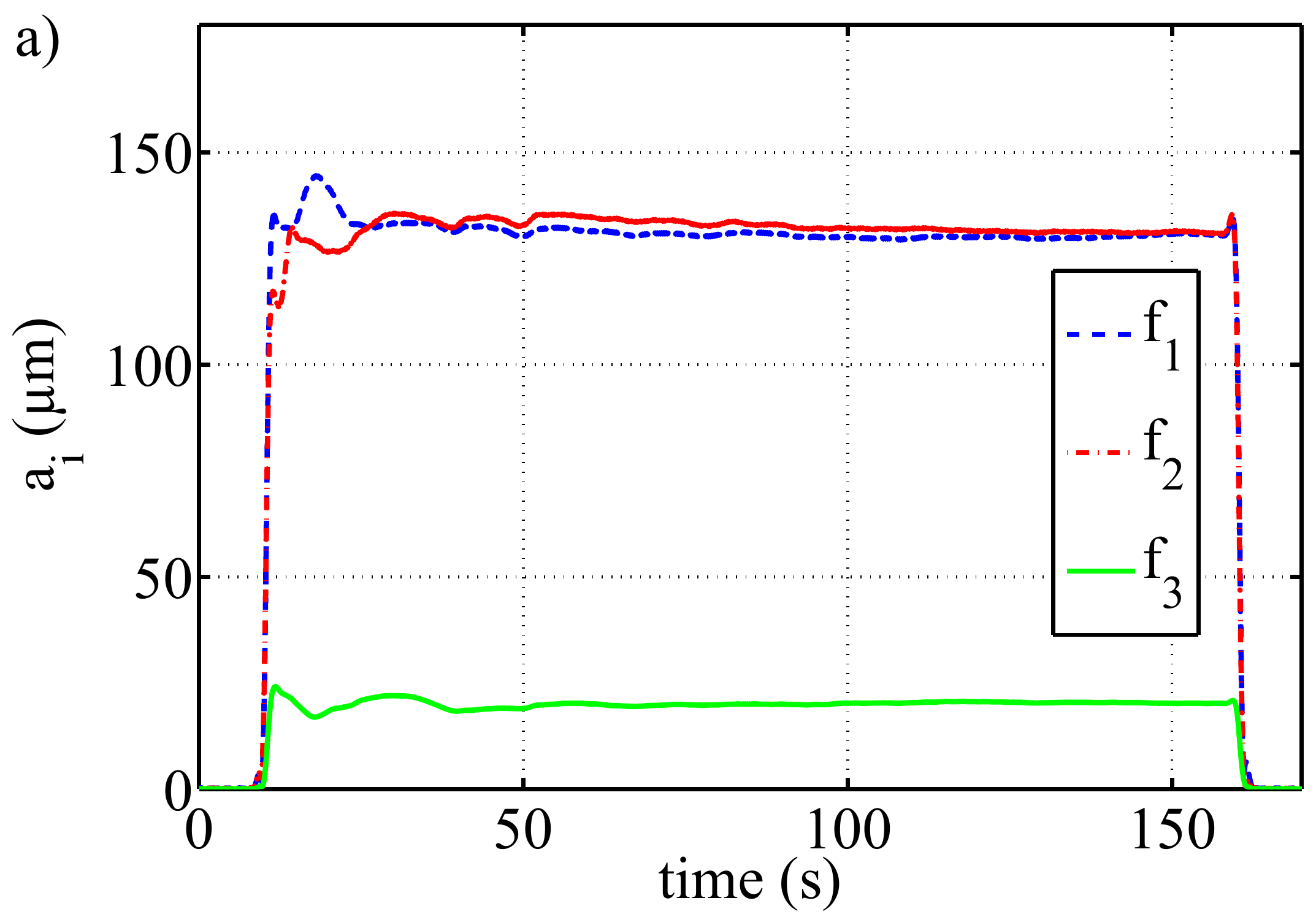} \hfill
\includegraphics[width=.85\columnwidth]{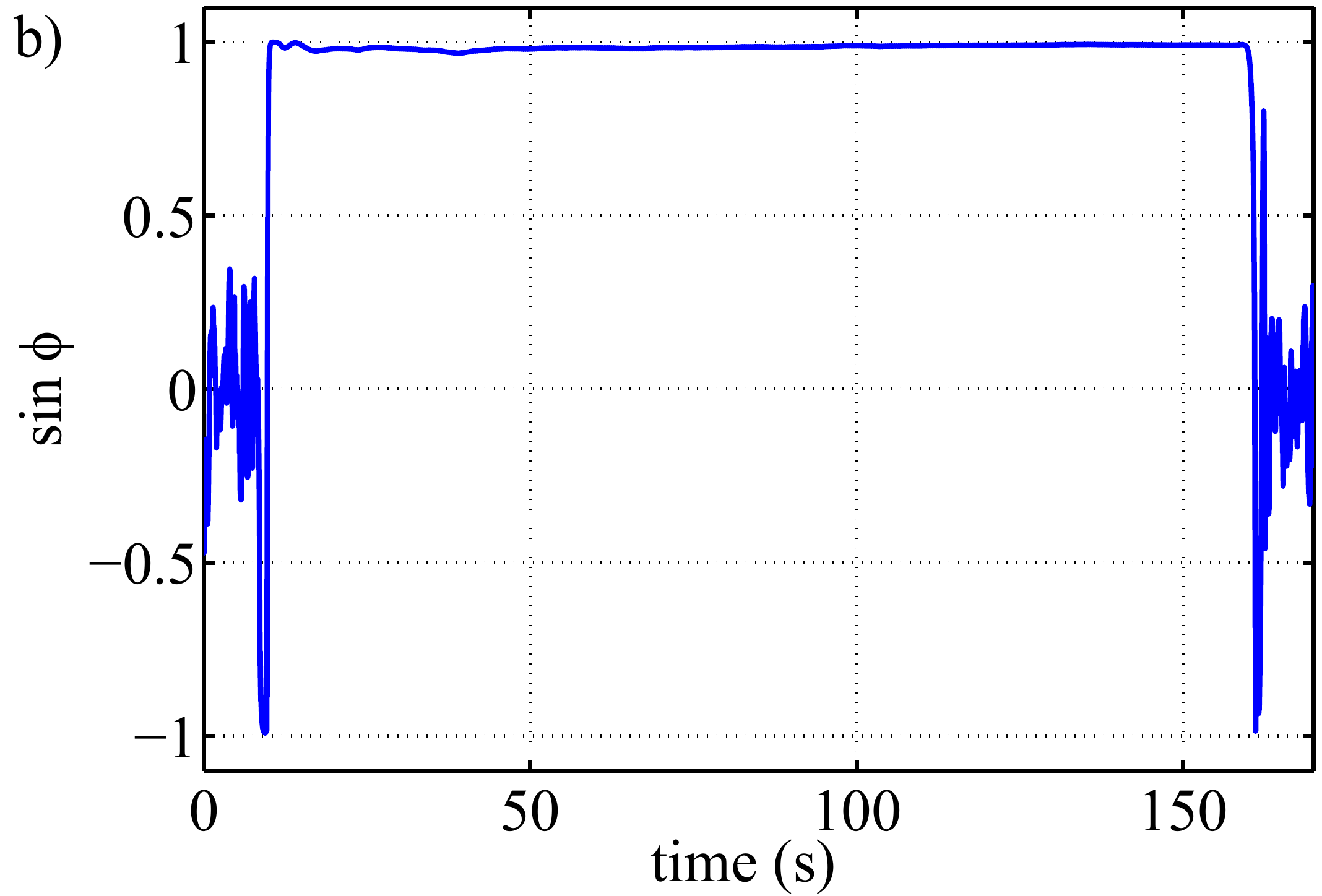}
\includegraphics[width=.85\columnwidth]{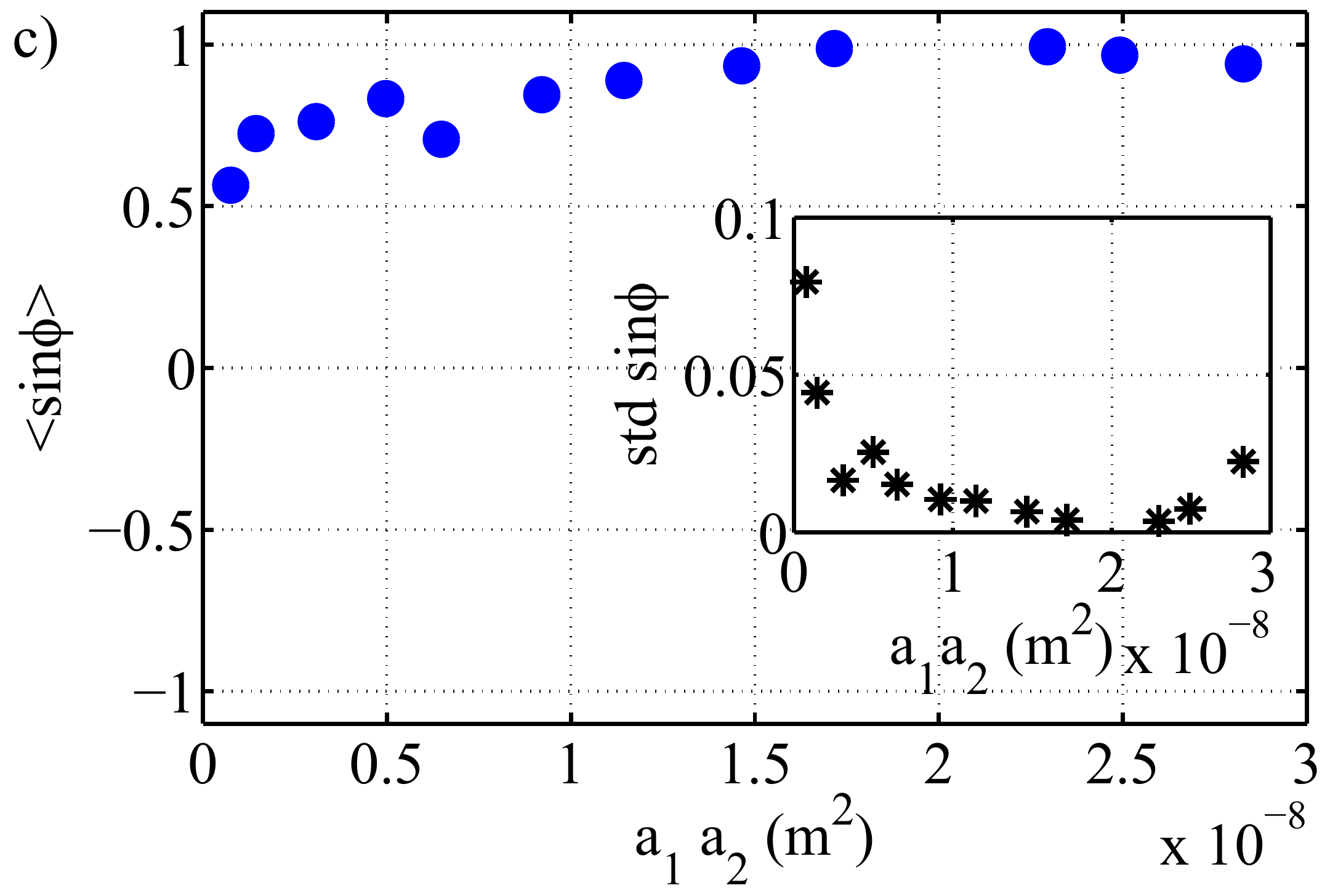}\hfill
\includegraphics[width=.83\columnwidth]{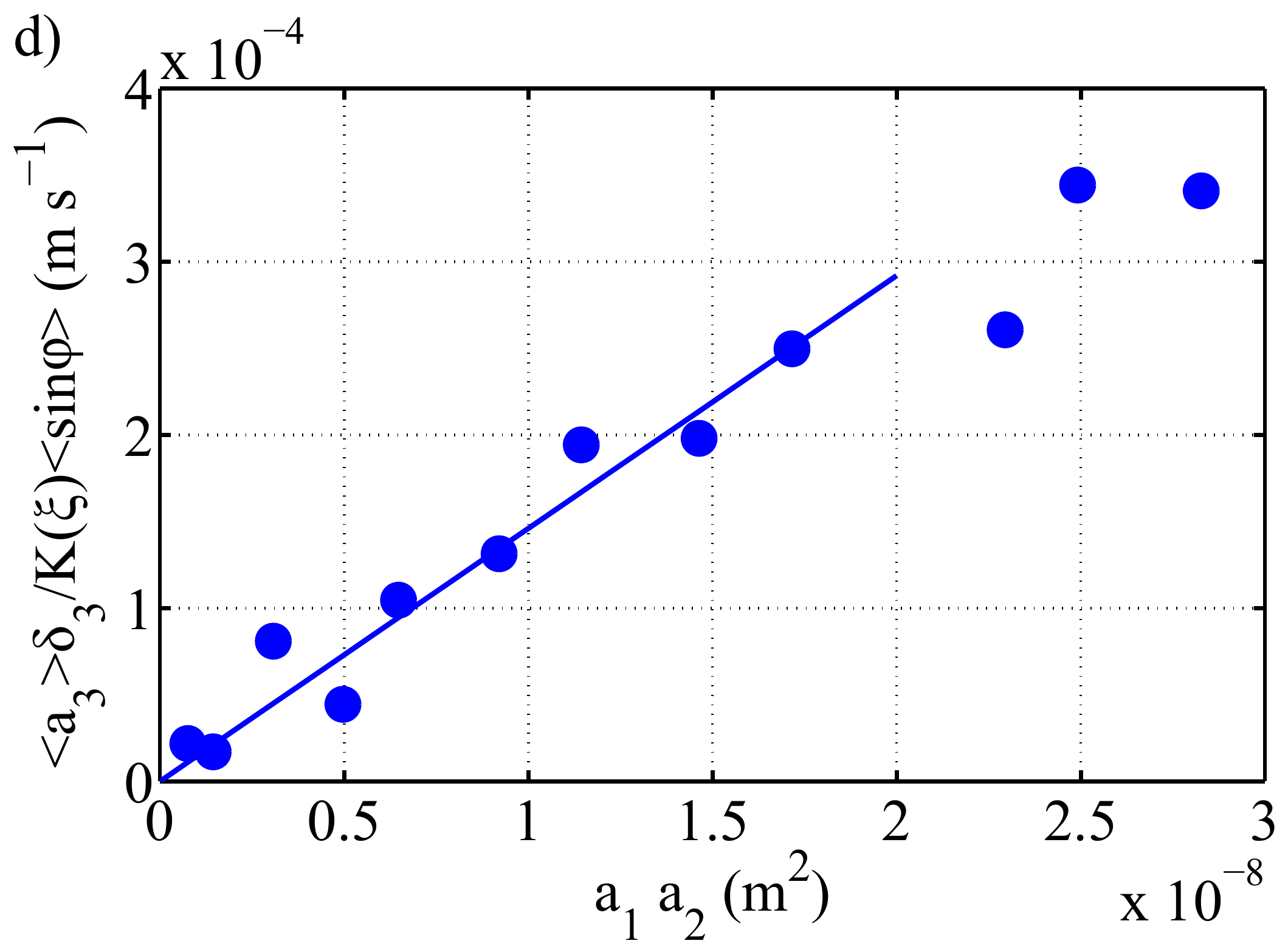} 
\caption{(color online) a) Amplitudes $a_i$ and b) $\sin\phi$ versus time for $f_1=15$ Hz, $f_2=18$ Hz and $\alpha_{12}=54$ deg. For these parameters, $a_1=130$ and $a_2=132$ $\mu$m. c)
Mean value of $\sin\phi$ versus $a_1a_2$ in stationary regime, with in the inset : the standard deviation of $\sin\phi$, quantifying fluctuations around the mean value. We observe that the phase locking occurs for the considered forcing amplitudes, but at low forcing $\sin\phi$ departs slightly from $1$ and fluctuations are stronger. d) In stationary regime $\langle a_3 \rangle \times\delta_3/(K(\xi)\,\langle \sin\phi \rangle)$ is found proportional to $a_1a_2$ for small and moderate forcing. Solid line is the linear fit valid for not too high values of $a_1a_2$.}
\label{sinphi_amplitudes_54_april}
\end{center}
\end{figure*}

After verifying the resonance conditions and showing that Eq.~\ref{a3} describes the spatial behavior of the daughter wave in stationary regime, we study now its temporal dynamics in more accurate measurements using the laser Doppler vibrometer. 12 realizations of the same experiment and for 12 increasing mother wave amplitudes have been performed.  Concerning the experimental protocol, at $t=10$\,s, excitation is started for 150 s and is then stopped at $t=160$ s, whereas wave heights are recorded during the total duration of the experiment lasting 170 s. \\
From the measurements, we examine the evolution of the amplitudes and the phases of the different components of the triad in the stationary regime at the position $(x_v,y_v)$. To do so, the vibrometer signal is filtered around each considered frequency $f_i\,$: the filter is of Butterworth type, of order 2 and with a passband of $2\Delta f$ ($\Delta f=0.5$ Hz). Resulting signals are then integrated over time, to transform vertical interface velocity into wave amplitudes. Then a Hilbert transformation is applied to the signal in order to extract separately the instantaneous wave amplitude $a_i(t)$ and the instantaneous phase $\psi_i= \mathbf{k_i} \cdot \mathbf{x}_v - \omega_i\,t + \phi_i$ of the different waves at the laser beam position $\mathbf{x}_v$.\\
Looking at the amplitudes $a_i$ averaged on 12 realizations in Fig.~\ref{sinphi_amplitudes_54_april} a), we can see that after a transient and a small overshoot, the amplitudes stabilize to a stationary value. At $t=160$\,s, when the excitation is stopped, the amplitudes return to zero. The growing and decaying transient regimes involve shaker dynamic response, wave propagation, viscous dissipation, and non-linear wave interaction. Due to the lack of temporal resolution resulting from the filtering operation, transients cannot be used to evaluate the growth coefficient $\gamma_3$ of the daughter wave. We notice also that the amplitude of the daughter wave, around 20 $\mu$m, is smaller than the amplitudes of the mother waves, around 130 $\mu$m ($a_3 <<\sqrt{a_1\,a_2}$). \\
Concerning the phase, Fig.~\ref{sinphi_amplitudes_54_april} b) displays $\sin\phi$ as a function of $t$. If the resonant conditions of Eq.~\ref{Res} are verified, by computing $ \psi_1+\psi_2-\psi_3$, we obtain the total phase $\phi=\phi_1+\phi_2-\phi_3$ and so the $\sin\phi$ term involved in Eq.~\ref{EqAmp}. During the transient regimes, phases $\phi_i$ are not related and $\sin \phi$ fluctuations are fast. Then once stationary regime is reached, $\sin \phi$ evolves slowly and stabilizes to $1$. Therefore there is a constant relation between the phases in the triad, \textit{i.e.} a phase locking around $\phi=\pi/2$ as expected theoretically to have a stationary phase behavior (see Eq.~\ref{EqPhase}). This constitutes a strong argument proving that the wave $3$ is created by the resonant interaction mechanism. \\
The influence of increasing the amplitudes of the two mother waves is presented in Fig.~\ref{sinphi_amplitudes_54_april} c) and d) for which both amplitudes $a_i$ and $\sin\phi$ are averaged first over time in the stationary regime (between $t=60$ and $140$\, s) and then on the 12 consecutive identical experimental realizations. In these experimental conditions, Fig.~\ref{sinphi_amplitudes_54_april} c) shows the mean value and the standard deviation of $\sin\phi$  versus $a_1a_2$. Except at the lowest amplitudes, standard deviation of $\sin\phi$ is quite small (below $0.02$), showing that a phase locking occurs. But we observe that the phase locking value differs at small amplitude from the expected value of $1$. Fig.~\ref{sinphi_amplitudes_54_april} d) presents the evolution of $a_3\delta_3/ (K(\xi)\sin \phi )$ versus $a_1a_2$. As expected from Eqs.~\ref{a3} and \ref{K}, we found a proportional behavior between the rescaled $a_3$ and $a_1a_2$, at least for not too high amplitudes. \\
Following Eq.~\ref{a3}, we estimate the interaction coefficient $\gamma_3$ for the daughter wave by computing the slope of the linear fit of $a_3\delta_3/( K(\xi) \sin\phi)$. To take into account the growth of the daughter wave on a distance $\xi$ between the beginning of the wave-train and the laser spot, the coefficient $K(\xi)$, has to  be evaluated. 
For these experiments the distance $\xi$ is roughly $30\,$mm, leading to $K(\xi)=0.33$.
The slope of the linear fit in Fig.~\ref{sinphi_amplitudes_54_april} d) is $4.85\, 10^{ 3}$ m$^{-1}.$s$^{-1}$, and gives thus an estimate of $\gamma_3$. We obtain thus an experimental estimation of the non-linear interaction coefficient $\gamma_{3exp}=1.46\,10^{ 4}$ m$^{-1}.$s$^{-1}$, which is $20 \%$ more than the theoretical value  $\gamma_3= 1.24\, 10^{ 4}$ m$^{-1}.$s$^{-1}$. Note that repeating the experiments with another triad (16, 23, 39) Hz leads to an estimation of the coefficient $\gamma_{3,exp}= 1.22\,  10^4$ m$^{-1}$s$^{-1}$, which is $13 \%$ less than the theoretical value $\gamma_{3,th}=1.41\, 10^4$\,m$^{-1}$s$^{-1}$.\\
These measurements validate thus the generation of a daughter wave from the interaction between two mother waves, for capillary-gravity waves in a closed tank. It is important to keep in mind that the wave-field presents inhomogeneity, due to two main reasons: first, the presence of a significant standing wave part, in addition to the main propagative part. Secondly the viscous dissipation decreases significantly the amplitudes of the mother waves as they propagate away from the wave makers. Moreover the frequencies $f_1$, $f_2$ and $f_3$ of the triad, are involved in other non-linear mechanisms. Nevertheless, the main features given by the resonant interaction theory are recovered in this experiment. We observe that the daughter wave verifies the resonant conditions in frequency and wave number. Furthermore, the total phase $\phi$ is locked to a value close to $\pi/2$. The non-linear interaction coefficient experimentally estimated is finally quite close to the theoretical one, considering that the experimental conditions are not strictly conform to the framework of the theory.

\section{Conclusions and discussion}
 We report an experimental study on three-wave interactions of capillary-gravity waves in a closed tank. We showed that the interaction between two mother wave trains at frequencies $f_1$ and $f_2 $ creates a daughter wave at $f_3=f_1+f_2$. For mother waves crossing at the resonant angle, we experimentally validate the three-wave resonant mechanism. By means of a spatio-temporal measurements, we verify that the spatial resonance condition is fulfilled. We also observe a phase locking, at $\phi=\pi/2$, between the different waves of the triad as theoretically expected . In the stationary regime, we measure the growth rate of the daughter wave amplitude $a_3$, which is found to be proportional to $\gamma_3\, a_1 a_2$, with $a_1$ and $a_2$ the mother wave amplitudes. Our quantitative estimation of the non-linear interaction coefficient $\gamma_3$ gives the correct order of magnitude with respect to the theoretical value, within an accuracy of less than $20\%$. This extensive study has been performed for two different triads (15, 18, 33) Hz or (16, 23, 39) Hz (the latter is not shown here). Similar results are found. This confirms that the features of three-wave interaction reported here can be generalized to different capillary-gravity triads within a frequency range of approximatively $10 < f < 50$\,Hz.\\
Several phenomena could be addressed to explain the slight departures between experimental and theoretical values: i) higher-order interactions at high forcing amplitude, ii) wave reflections on the boundary of the tank producing standing waves, that reduce homogeneity of the wave-field where the daughter wave is studied, iii) issue to localize the beginning of the daughter wave in local measurements, and iv) finite width of the wave-trains that could also modify the derivation of interaction coefficients~\cite{Bourget2014}.\\  
The model used here, (Eqs.~\ref{EqAmp}, \ref{EqPhase} and \ref{a3}), involving viscous dissipation as a perturbation~\cite{Henderson1987_1,McGoldrick1970,Craik}, describes appropriately our results in the stationary regime. The characteristic non-linear time $\tau_{nl}$ for the growth of $a_3$ can be approximated as $1/\tau_{nl}=(\gamma_3 \, \sqrt{a_1 a_2})$. For the triad $(15,18,33)$\,Hz, $1/\tau_{nl} \approx 1.24$\,s$^{-1}$ taking ${a_1 a_2}\approx1.10^{-8}$~m$^2$ (Fig.\ref{sinphi_amplitudes_54_april} d), leading to $\tau_{nl}\approx0.81$~s. The dissipative time is evaluated as $\tau_{d}=1/\delta_3=0.23$\,s. For capillary waves, the dissipation is thus always too strong to be neglected and it is surprising that inviscid theories provide correct values for the interaction coefficients. Consequently a theoretical effort describing three-wave resonant interactions from Navier-Stokes equations for surface waves, would be of prime interest.\\
Note that such an extensive experimental study of three-wave interaction for gravity-capillary waves on the surface of a fluid has never been tested, although this process transfers energy at small scale in wave turbulence. Indeed, previous works using only collinear wave-trains, investigated the degenerated case of Wilton ripples~\cite{McGoldrick1970,Kim1971} or the subharmonic generation where one wave at high frequency produces two waves at a lower frequency~\cite{Banerjee1982,Hogan1984,Henderson1987_1}. 
Here, we emphasized that, for a given couple of mother waves frequencies, the resonance conditions impose, theoretically from Eq.~\ref{Eqalpha}, the value of the resonant angle between the two mother waves. When the angle is experimentally fixed to the resonant angle, we have shown here that three-wave interactions are correctly described by the resonant interaction theory. As these interactions are the elemental mechanism of capillary wave turbulence, this study seems to show that resonant interactions insure energy transfer through the scales. The average flux $\epsilon_3$ transferred to the daughter wave by unit of area and density, can be estimated from the energy the wave $3$~\footnote{The energy per surface unit of each triad component can be expressed as $E_i(t)=\dfrac{1}{2} \rho \omega_i 
\,v_{pi}\, {{a}^2}_i (t)\,$, with $v_{pi}$ the phase velocity of the wave $i$~\cite{Simmons1969}}.  After spatial and temporal averaging, we find $\left\langle \epsilon_3 \right\rangle \approx 2 \,10^{-7}$m$^3.$s$^{-3}$, which is close to values obtained in capillary wave turbulence experiments in water, when energy flux is evaluated through the dissipated power~\cite{Deike2014}. Nevertheless it is important to evaluate and understand the relative contribution of three-wave quasi-resonances [36] and non-resonant interactions [6] in front the resonant interactions for gravity-capillary wave experiments in laboratory.

\section*{Acknowledgments}
The authors thank the ANR Turbulon 12-BS04-0005 which funded this work. L. D. was supported by grant from ONR and NSF to W. K. Melville. We acknowledge also C. Laroche for technical assistance.

\appendix*

\section*{Appendix : Attenuation of the waves due to viscous damping}
\label{Att}
\noindent
In our experiments, the amplitude of a daughter wave is due to a balance between viscous attenuation and three-wave interactions. In order to experimentally determine the viscous dissipation or attenuation coefficients $\delta_i$, we have performed experiments with only one wave maker generating a monochromatic wave, with $f_1$, $f_2$ and $f_3$ generated separately. The amplitude of the wave has been recorded every centimeter with the laser vibrometer put on a linear translation stage. A beach with a slope around $45$ deg and lateral walls have been placed in the tank in order to avoid reflections of the waves when encountering the solid curved walls of the tank. 

\begin{figure}
\begin{center}
\includegraphics[width=0.85\columnwidth]{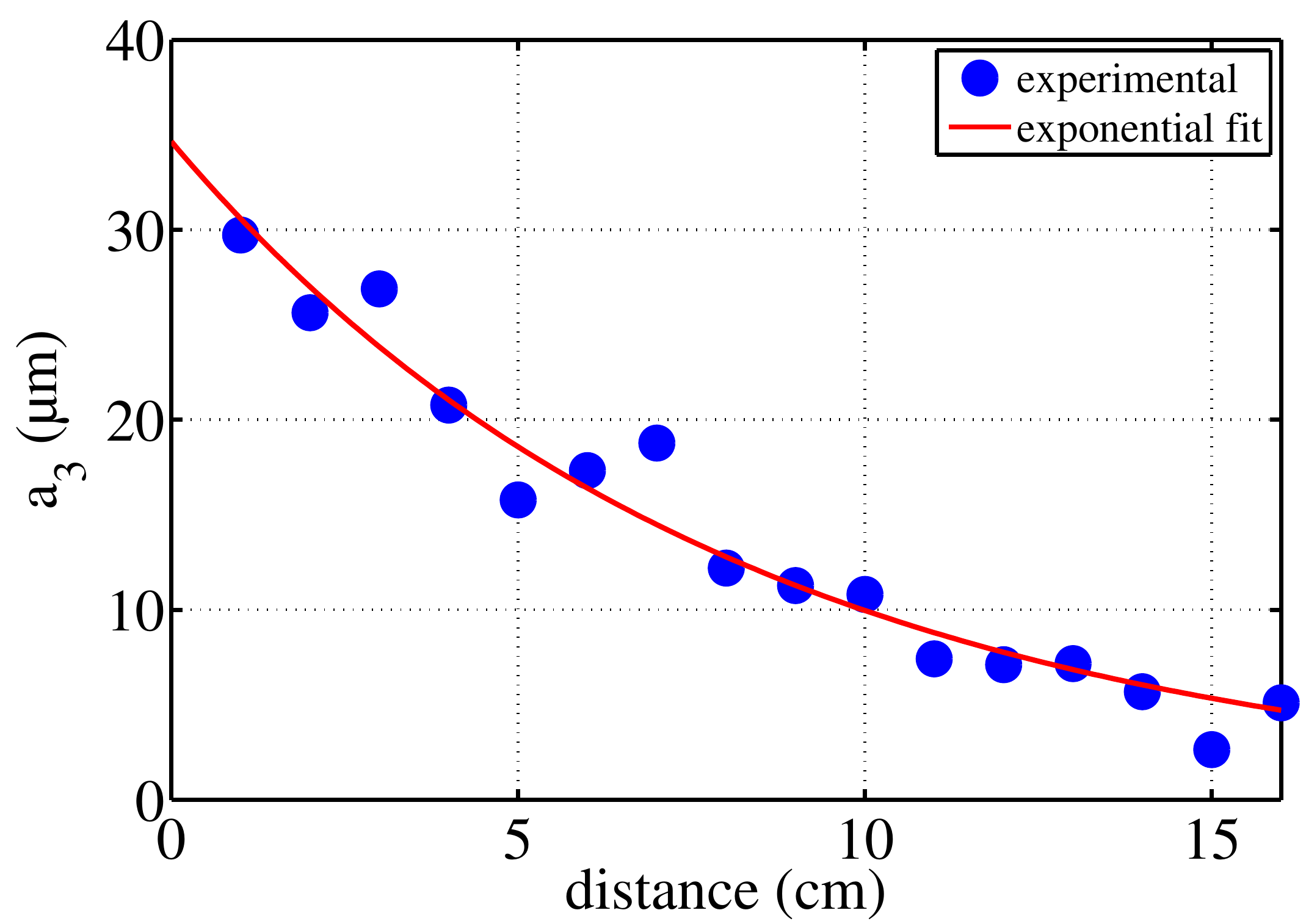}
\caption{(color online) Decrease of the amplitude of the wave with frequency $f=33$\,Hz versus the distance from the wave maker, for a small excitation forcing amplitude.}\label{attenuation_33}
\end{center}
\end{figure}
The evolution of amplitude is expected to decrease exponentially with the distance from the wave maker. For a given frequency, the attenuation length $l_i$ can be estimated from the decaying exponent of an exponential fit performed on the experimental amplitudes as shown in Fig.~\ref{attenuation_33}, for a frequency equal to $33$\,Hz. Then the coefficients $\delta_i$ are simply computed by $\delta_i=v_{gi}/l_i$. The group velocities are calculated using $v_{gi}=\partial{\omega_i}/{\partial k_i}$ for deep water dispersion relation, $\omega_i=\sqrt{g k_i+\frac{\sigma}{\rho} k_i ^3}$. This leads to:
\begin{equation}
\partial{\omega_i}/{\partial k_i}=\frac{1}{2}\left( g+\frac{3 \sigma}{\rho} k_i^2\right) \left( gk_i+\frac{\sigma}{\rho} k_i^3\right) ^{-1/2}
\end{equation}
\noindent
Despite the experimental care, in preparing the solution of \chemform{TiO_2}, the dissipation exponents are found after one hour, well described by the inextensible free-surface model due to the presence of insoluble surfactants~\cite{Lamb1932,VanDorn,Deike2012,HendersonSegur2013}. The values of the viscous dissipation coefficients are thus computed using the formula $\delta_{i\,{model}}=\sqrt{2} \sqrt{\nu \omega_i} k_i /4$. The values of the norms of the wavevectors $k_i$ (found with the graphical resolution of Fig.~\ref{triad_graphs1} b) with $\sigma=60$\,mN/m), the phase and group velocities $v_{pi}$ and $v_{gi}$, the attenuation parameters $\delta_{i\,{model}}$, $l_i$ and $\delta_i$ are reported in Tables~\ref{table2} and \ref{table3}. Note that the measured coefficients are close to the value of $\delta_{i\,{model}}$ (maximum relative error around equal to 7\%).


\small

\begin{table}[h!]
\begin{minipage}[t]{.65\linewidth}

\begin{tabular}{|l|c|c|c|c|c|r|}
 \hline
&$k_i$ (m$^{-1})$&$v_{pi}$ (m. s$^{-1})$&$v_{gi}$ (m. s$^{-1}$) \\
 \hline
$f_1=15$ Hz & 428 &0.220&0.227
\\
 \hline
$f_2=18$ Hz & 507 &0.223&0.248
\\
 \hline
$f_3=33$ Hz & 834 &0.249 &0.326
\\
 \hline
 \hline
$f_1=16$ Hz & 455 &0.221&0.234
\\
 \hline
$f_2=23$ Hz & 626 &0.231&0.278
\\
 \hline
$f_3=39$ Hz& 946 &0.259 &0.349
\\
 \hline
\end{tabular}
\end{minipage}
\caption{Norms of the wavevectors, phase and group velocities. The different values are calculated with $\sigma=60$ mN/m and $\rho=1000$\,kg.m$^{-3}$.}
\label{table2}
\end{table}
\normalsize

\small

\begin{table}[h!]
\begin{minipage}[t]{.65\linewidth}

\begin{tabular}{|l|c|c|c|c|c|r|}
 \hline
&$\delta_{i\,model}$ (s$^{-1}$) & $l_i$ (m)&$\delta_i$ (s$^{-1}$) \\
 \hline
$f_1=15$ Hz&1.49&0.162&1.40
\\
 \hline
$f_2=18$ Hz &1.93&0.128&1.94
\\
 \hline
$f_3=33$ Hz&4.31&0.080&4.07
\\
 \hline
 \hline
$f_1=16$ Hz&1.63&0.122&1.92
\\
 \hline
$f_2=23$ Hz &2.70&0.103&2.70
\\
 \hline
$f_3=39$ Hz &5.31&0.061&5.71
\\
 \hline
\end{tabular}
\end{minipage}
\caption{Viscous damping coefficients (viscous model :  $\delta_{i\,{model}}=\sqrt{2} \sqrt{\nu\omega_i} k_i/4$), experimentally measured attenuation lengths and corresponding deduced viscous damping coefficients. The different values are calculated with $\sigma=60$\,mN/m.}
\label{table3}
\end{table}
\normalsize

Note as previously mentioned, that gravity-capillary waves are sensitive to the contamination of the surface that is an unavoidable effect. In order to check if contaminants play a significant role in the present experimental conditions, we have performed for an arbitrary angle between the two wave makers, a complete series of experiments with a home-build plastic cover and the same series another day without the plastic cover. Results appear to be comparable with and without cover, excepted during the first hour.

Finally viscous damping coefficient have been also measured for the solution of Intralipids using the DLP method. Spatial decay of sinusoidal wave-trains are found compatible with the inextensible free-surface model, $\delta_{i\,{model}}=\sqrt{2} \sqrt{\nu\omega_i} k_i /4$. Compared to the solution of \chemform{TiO_2}, this solution has a lower surface tension $\gamma=55$\, mN.m$^{-1}$ and a slightly higher viscosity $\nu=1.24~$m.s$^{-2}$. We obtain thus for example for $f_3=33\,$Hz, $v_{g3}=0.316\,$m.s$^{-1}$ and $\delta_3=4.83$\,s$^{-1}$. Therefore both liquids used in this experimental study have analogous physical properties and we observe the same behavior for the three-wave resonance mechanism.

\bibliography{threewaves_20160111}

\end{document}